\PassOptionsToPackage{hyphens}{url}

\documentclass[11pt,a4paper]{article}
\pdfoutput=1

\usepackage{fullpage}
\usepackage{setspace}
\usepackage{amsmath}
\usepackage{amsfonts}
\usepackage{amssymb}
\usepackage{fix-cm}
\usepackage[toc]{appendix}
\usepackage{xcolor} 
\usepackage{hyperref}
\hypersetup{colorlinks, allcolors=blue}
\usepackage{tikz}
\usepackage[numbers]{natbib}
\usepackage[most]{tcolorbox}
\usepackage{booktabs}
\usepackage{xcolor}
\usepackage[UKenglish]{babel}

\begin{document}
\begin{titlepage}

\centering

{\huge\bfseries{Superconducting Microwave Detector Technology for Ultra-Light Dark Matter Haloscopes and other Fundamental Physics Experiments: Device Physics (Part II)}\par}

\vspace{0.5cm}
{ David J. Goldie $^1$, Stafford Withington$^2$ and Christopher N. Thomas$^1$ \par}

\vspace{0.5cm}
{$^1$\emph{Cavendish Laboratory, University of Cambridge, JJ Thomson Avenue, Cambridge, CB3 0HE, UK}\par}
{$^2$\emph{Clarendon Laboratory, University of Oxford, Parks Road, Oxford, OX1 3KU, UK}\par}
\vspace{0.5cm}
\today

\begin{abstract}
\noindent We consider and compare candidate  superconducting detector technologies that might be applied to the
readout of cavity-axion haloscopes and similar fundamental physics experiments. 
We conclude that a transition edge sensor (TES) configured with ballistic-phonon thermal isolation operated with a superconducting transition temperature of order $30\,{\rm mK}$ would provide quantum-limited detection performance at frequencies above $5\,{\rm GHz}$. 
This would permit the realisation of integrated homodyne detectors based on TESs that we believe would make a unique contribution to a variety of fundamental physics experiments, particularly those based on reading out microwave cavities in the quantum ground state.
\end{abstract}

\end{titlepage}

\newpage

\tableofcontents

\newpage

\onehalfspacing

\section{Introduction}\label{sec:introduction}

In a companion paper, \cite{Chris2023} we developed a comprehensive theoretical description of ultra-low-noise microwave and millimetre-wave homodyne detection. Homodyne detection has been used extensively at optical wavelengths in quantum-optics experiments,\cite{Optics_Shaked,Optics_Lvovsky}, and even finds important use on large-scale experiments such as the gravitational wave interferometer LIGO \cite{LIGO_Fritschel}. We are not aware, however, of the homodyne method being applied at microwave and millimetre wavelengths using ultra-low-noise superconducting detectors. A wide range of potential applications exist; for example, in the search for ultra-light dark matter, such as axions and dark photons, using haloscopes and other innovative schemes. At microwave and millimetre wavelengths, however, the primary detectors operate in the classical-to-quantum transition, which makes theoretical analysis more involved than simply assuming photon-counting statistics. At these wavelengths, classical and photon-counting noise are present in similar measure, and so the trade off between the two becomes important. In addition,  the classical and quantum noise of suitable devices makes a significant contribution to the overall noise budget, and must be woven into the theoretical models in an appropriate way. Crucially, the operation of homodyne detection must be described in a way that is accessible by the traditional RF solid-state devices community. 

The analysis in our companion paper is device agnostic and so the question remains as to which superconducting detector technologies are most-appropriate for the application. The questions go well beyond the matter of noise performance, but are critically related to considerations such as tuning range, instantaneous  readout (baseband) bandwidth and saturation power. Achieving ultra-low noise operation, whilst ensuring linearity up to high power, as needed by the pump, depends on the specific device used. In this report, we consider a range of different superconducting detectors, and assess which, if any, are suitable for realising integrated microwave-to-submillimetre-wave homodyne detectors.

The paper is structured as follows: Section~\ref{sec:requirements} describes the experimental arrangement and considers bolometer requirements for homodyne detection. We define a simple figure-of-merit (FoM) for detector performance. Section~\ref{sec:Device_types} provides an assessment of a number of cryogenic detectors, which we identified as being potentially interesting for homodyne applications. Subsequently, we found that {\it any} bolometric detector operating close-to thermodynamic limits would be suitable, the choice then depends on readout mode (particularly if frequency modulation is inherent to the readout) and the maturity of the technology. Section~\ref{sec:Thermal_inputs} discusses the inputs used for thermal modelling. Section~\ref{sec:FoM} discusses the figures-of-merit of our candidate bolometers. Section~\ref{sec:TES_for_homodyne} shows results of thermal modelling of our own preferred device, the Transition Edge Sensor (TES), in the context of its usefulness for realising integrated homodyne detectors. Section~\ref{sec:HEB_performance} presents thermal modelling and the suitability of Hot Electron Bolometers (HEB) for homodyne detection. Bolometric devices require superconductors having suitably low $T_c$ and controllable materials for operation in the range $10-50\,{\rm mK}$, and so in Section~\ref{sec:Tc_model_measure} we describe modelling of proximity effects in mulitlayers, and measurement of transition temperature $T_c$ for thin-film multilayers fabricated from Ti and Au. We found that low $(< 50 mK)$ transition temperatures are indeed possible,  and predict the bilayer thicknesses needed for the $30\,{\rm mK}$ operation of either TESs or HEBs. In Section~\ref{sec:Conclusions} we summarise and conclude our findings.

\section{Review of Requirements}\label{sec:requirements}
\subsection{Homodyne Detection}\label{sec:Homodyne_detection}

\begin{figure}
\centering
\includegraphics[width=15cm]{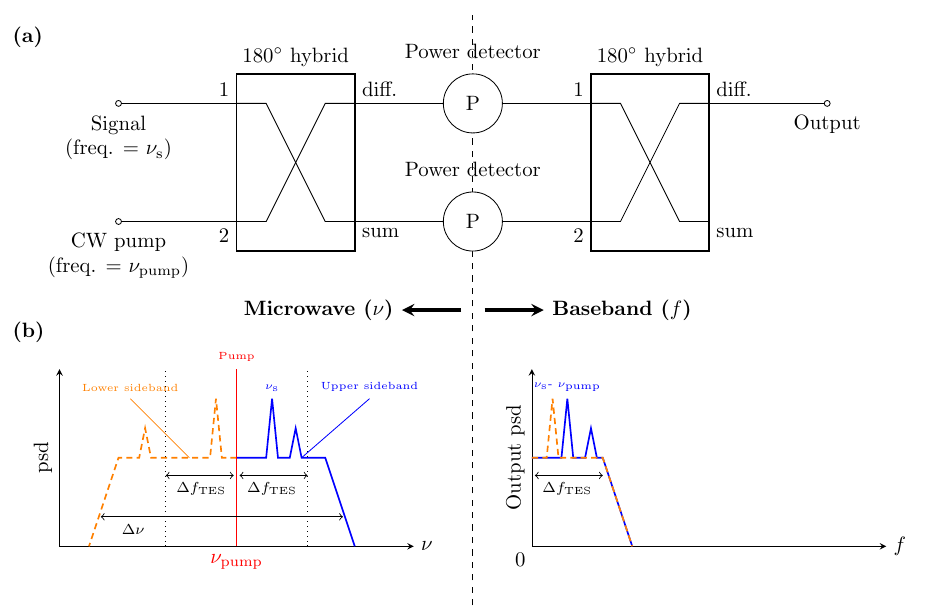}
\caption{\label{fig:homodyne_concept} Schematic showing the homodyne power detection concept. Matched power detectors and two RF hybrids  down-convert the microwave signal of interest at frequency $\nu_s$ and pump signal at $\nu_{pump}$ into the baseband measurement frequency range $f$ and output signal for further analysis.}
\end{figure}

Our aim to to realise integrated homodyne  detectors operating in the range  $\nu_s=1-20\,{\rm GHz}$.\cite{backes2021quantum} Figures~\ref{fig:homodyne_concept}(a) and (b) illustrate the homodyne concept. An unknown signal at frequency  $\nu_s$ is mixed with a pump tone at frequency $\nu_{pump}$ and {\it total}  power $2P_{pump}$. In this note we define $P_{pump}$ as the power applied to {\it a single} power detector. Both $\nu_{pump}$ and  $P_{pump}$ can be varied easily. An RF hybrid outputs the sum and difference of the signal and pump and feeds them into two matched power detectors.  We assume that the detectors are sensitive to the powers applied to their inputs. The outputs of the detectors only respond to changes in detected power that occur within the output bandwidth $\Delta f$. The signal of interest is then the difference of the detector outputs, at frequency $f=\lvert \nu_s-\nu_{pump}\rvert$. In this paper, we use $f$ to denote baseband  (or equivalently output) frequencies. The required signal, at baseband, then appears as excess structure in the power spectral density (${\rm PSD\,}(f)$) of the overall output, as shown in Fig.~\ref{fig:homodyne_concept}(b). 

\subsubsection{Homodyne Detection Bandwidth Requirement}\label{sec:operation_bandwidth}

In order to achieve best performance we would like to  ensure that the bandwidth available for the down-converted signal  
$\Delta B$ is greater than the axion linewidth, so that the detector operates as a matched filter or that an array of matched filters can be synthesised digitally  at baseband. The linewidths of halo axions are expected to be of order $500-1000\,{\rm Hz}$ \cite{Daw1998a}, and we take this as the minimum requirement for the output bandwidth. For a TES bolometer $\Delta B\sim f_{\rm 3 dB}= 1/ (2\pi\tau_{eff})$ and the approximation becomes an equality if detector noise is determined by thermal noise associated with heat flow to the bath. $\tau_{eff}$ is the effective detector response time, which can be shorter than the thermal response time $\tau_0$ because of the presence of electrothermal feedback (ETF). We assume throughout  that $1/f$ noise in the readout chain is sufficiently low that it does not eat appreciably into the output bandwidth available.  Also, that the $1/f$ noise in the region of the output bandwidth occupied by the down-converted axion signal is sufficiently small that it can be ignored. These are both acceptable assumptions.

\subsection{Noise temperatures for Homodyne Detection}\label{Noise_temperatures}

This Section summarises detector requirements for homodyne detection. More details can be found in Ref.~\cite{Chris2023}.
The effective receiver noise temperature is given by 
\begin{equation}\label{eqn:T_rcvr}
	T_{rcvr}= \frac {{\rm NEP}^2} {2 k_B P_{pump}},
\end{equation}
where ${\rm NEP}$ is the input-referenced detector Noise Equivalent Power,  $P_{pump}$ is the pump power for a {\it single} power detector as shown in Fig.~1 (a),
and  $k_B$ is Boltzmann's constant.  The quantum-limited noise temperature $T_q$ is given by
\begin{equation}\label{eqn:T_q}
	T_{q}= \frac{h\nu} {2 k_B},
\end{equation}
where $\nu$ is the detection frequency and $h$ is Planck's constant.  
Limiting performance is achieved when $T_{rcvr}\ll 2T_{q}$.

For a general bolometer, the power flow $P_{bath}$  between the bolometer at temperature $T_c$ and the heat bath at temperature $T_b$ can be parameterized such that 
\begin{equation}\label{eqn:P_bath}
	P_{b}=K_{bath}(T_c^n-T_b^n),
\end{equation}
where $K_{bath}$ and $n$ are constants that depend on the nature and/or dimensions of the structure or mechanism providing the  thermal isolation.  
The pump tone $\nu_p$ is well-outside the thermal or electrical response frequencies  of the bolometer so that the bolometer only responds to the time-averaged incident power. The expected signal power is also small. This means that in order to retain the beneficial effects of ETF  a  TES, for example, would also need to be DC biassed in addition to the pump plus signal powers. 
For a TES with an initial operating point $R=R_n/4$ (i.e. DC-biassed, without $P_{pump}$),  we would expect that applying $P_{pump}=\kappa_{pump}P_{b}$ with $\kappa_{pump}= 0.5$ would increase the TES to an operating resistance $R_0\sim R_n/2$, close-to the maximum value of
$\alpha$ for many observed TES resistive transitions. For later modelling we assume  $\kappa_{pump}=0.5$ in order to compare bolometer performance.

Combining terms we can introduce a dimensionless  figure-of-merit (FoM) for homodyne detection, which we aim to {\it maximise},  such that
\begin{equation}\label{eqn:FoM}
	{\rm FoM}=\frac {2T_{q}} {T_{rcvr}} = \frac {2 h\nu \kappa_{rf} \kappa_{pump} P_{b}}
	 {{\rm NEP}^2},
\end{equation}
where we have included an additional RF coupling efficiency $\kappa_{rf}$. We will use Eq.~\ref{eqn:FoM} later to compare different power detectors. 

For a  thermal detector the differential conductance between the active film and the heat bath   is $G_b=dP_b/dT_c$, which from Eq.~\ref{eqn:P_bath} is given by
\begin{equation}\label{eqn:G_bath}
 G_b=nK_{bath}T_c^{n-1}.
 \end{equation}
 $G_b$ determines the thermodynamic limit for the achievable NEP:
\begin{equation}\label{eqn:NEP}
	{\rm NEP}=\sqrt{4k_B\gamma G_b T_c^{2}},
\end{equation}
where $\gamma$ is a further factor introduced to take account of the temperature difference between the active film and the heat bath. Modelling suggests $\gamma\sim0.7$ depending on the details of the heat flow, but, unless stated and for later modelling, we assume that $\gamma=0.75$. Practical considerations, such as read-out noise, also need to be accounted for in the realised NEP.

Combining Eqs.~\ref{eqn:FoM}, \ref{eqn:G_bath} and \ref{eqn:NEP} we can refine the FoM for bolometric detectors so that
\begin{equation}\label{eqn:FoM2}
	{\rm FoM}= \frac { h\nu \kappa_{rf} \kappa_{pump} \phi(T_c,T_b)}     {2 k_B  T_c \gamma  },
\end{equation}
where $\phi(T_c,T_b)=\left( 1-\left( T_b/T_c \right)^n  \right)$. $\phi(T_c,T_b)\sim 1$ provided $T_b\lesssim T_c/2$ and higher values of $n$ are preferred. We use this equation later to compare  detector technologies. 
A FoM $\geq 1$ indicates quantum-limited detection.

\subsection{Figure of Merit}\label{sec:FoM_comments}

Reviewing Eqs.~\ref{eqn:FoM} and \ref{eqn:FoM2} when using bolometers for homodyne detection we see the following:

\begin{enumerate}
\item The detector NEP should be minimised.
\item  $\kappa_{pump}$ should be maximised.
\item $P_{pump}$ should be maximised. Although the acceptable pump power depends on detector characteristics. 
\item The trade-off between $P_{pump}$ and NEP is parameterized in Eq.~\ref{eqn:FoM2}.
\item Operating temperature and bath temperature should be minimised.
\item Lower values of  $n$, the power-flow index, are preferred.
\item Quantum-limited performance becomes more difficult as the operating frequency is reduced. 
\end{enumerate}

\section{Assessment of Device Types}\label{sec:Device_types}

In this section, we summarise the attributes of the main competing technologies. It is important to distinguish detectors by class: In Sec.~\ref{sec:Will_detectors_with_gap_work} we consider superconducting detectors having an energy threshold, whereas in Secs.~\ref{sec:TES} to \ref{sec:S1IS2_detectors} we consider bolometric detectors having no energy threshold.
 
\subsubsection{ Detectors having  an energy detection threshold}\label{sec:Will_detectors_with_gap_work}

At first sight there is no clear reason why a  superconducting detector operating  at  temperature $T_b$ well-below its
superconducting transition temperature $T_c$ could not be used as a power detector in the homodyne scheme.
For optimal performance this type of device is usually operated at $T_b\lesssim 0.2 T_c$. If $T_b\ll T_c$, a BCS superconductor has an energy gap  $\Delta=1.76 k_b T_c$. The minimum incident photon energy required to break a Cooper pair is $h\nu_{\min} = 2\Delta$.  Superconducting detectors having an energy gap have very low efficiency for recording sub-gap quanta. Detecting the frequencies of most interest in an axion haloscope appears to be exceedingly challenging for this type of device.
 
If we want to detect photons using a pair-breaking detector down-to a  frequency $\nu_s=\nu_0$ then we find $T_c=13\,{\rm  mK/GHz}\times \nu_0$ and $T_b=2.6\,{\rm  mK/GHz}\times \nu_0$. This implies for example $T_c=70\,{\rm mK}$ and $T_b =14\,{\rm mK}$ for  a detector relying on pair-breaking engineered to have detection {\it sensitivity}  at $\nu_0=5\,{\rm GHz}$. There is  a  further consideration. Photon attenuation in a superconductor, even above its pair-breaking threshold $2\Delta$, is non-linear (see for example de~Visser\,\cite{deVisser2015non} or 
Tinkham\cite{tinkhamintroduction}). The ratio of superconducting to normal-state attenuation has the limit $ \left( \alpha_s/\alpha_N\right) \vert_{h\nu^+\to 2\Delta}=0$ making calibration considerably more difficult close-to the detection threshold. In addition, few elemental superconductors are available with the required low $T_c$,  although in principle  these can be engineered using superconducting-normal metal bilayers (or multilayers) \cite{zhao2017exploring,zhao2018calculation}.
 
These considerations do not rule-out the possibility of upgrading existing ultra-low-noise pair-breaking detectors for homodyne detection over some of the frequency range of interest, but suggest that a considerable technical effort would be needed to make the approach work. It seems, therefore, that a number of very low-noise, low temperature detectors are not suitable for the application: for example, this excludes Kinetic Inductance Detectors, Quantum Capacitance Detectors, and more. We  will discuss separately the possibility of using thermal kinetic inductance detectors (TKIDs),  which do not rely on pair-breaking, in Sec.~\ref{sec:TKIDs}, and we consider asymmetric STJs in Sec.~\ref{sec:S1IS2_detectors}. 

\subsection{Transition Edge Sensors}\label{sec:TES}

Transition edge sensors (TES) are a mature cryogenic technology that, when operated at $T_c\sim 100\,{\rm mK}$, typically give an NEP of $3-5\times10^{-19}\,{\rm W/\sqrt{Hz}}$ along with saturation powers of order $P_{sat}\sim 5-10\,{\rm fW}$ or higher. An ultra-low-noise TES is usually engineered to operate within its superconducting-normal resistive transition. The TES is voltage biased and read out using a DC SQUID \cite{Irwin2005}. The SQUID current noise and the low operating temperature (hence low TES current noise) place their own constraints on the maximum normal state and operating resistances $R_n$ and $R_0$ respectively. We comment on this requirement in Sec.~\ref{sec:Rn_TES_max}. Electrothermal feedback in the TES reduces the response time $\tau_{eff}$ such that $\tau_{eff}=\tau_0/(1+\alpha/n)$ for $\alpha\gg 1$ \cite{Irwin2005}, where $\tau_0=C/G_b$ is the intrinsic response time. Here, $C$ is the total heat capacity of the device, and $G_b$ characterizes the thermal conductance to the heat bath. $\alpha=\left(T/R  \right) dR/dT$ parameterizes the sharpness of the S-N resistive transition as a function of temperature.  For a sharp transition $\alpha$ can be large $( \sim 100)$, enhancing appreciably the readout bandwidth. The thermal conductance to the heat bath $G_b$ can be engineered to behave diffusively or ballistically~\ref{sec:Diffusive_phonon_isolation}, \ref{sec:Ballistic_isolation}
or even through electron-phonon decoupling~\ref{sec:e_ph}. The operation of electron-phonon decoupled TESs seems to have been first demonstrated by Cabrera {\it et al.}\cite{cabrera1998detection}. We discuss electron-phonon decoupled TESs in Sec.~\ref{sec:HEB} as they are often described as hot electron bolometers (HEBs). 

\subsection{Hot Electron Bolometers}\label{sec:HEB}

We must immediately distinguish between diffusion-cooled HEBs (Sec.~\ref{sec:Diffusion_cooled_HEB}), designed for high frequency detection and mostly needing a high operating temperature, and low temperature nano-HEBs (Sec.~\ref{sec:Low_temp_HEB}), designed to minimise the NEP. The latter appears most suitable for ultra-low-noise homodyne detection schemes.

\subsubsection{Diffusion-Cooled Hot Electron Bolometers}\label{sec:Diffusion_cooled_HEB}

In a diffusion-cooled HEB the absorber is engineered to maximise the cooling rate of the electrons and hence baseband bandwidth. These devices are typically operated at $T_b\sim 4\,{\rm K}$ and are often used as mixers for high-frequency ($\nu_s\gtrsim 1 \,{\rm THz}$) photon detection. They comprise a thin, narrow resistive film of NbN or NbTiN with $T_c>12\,{\rm K}$, and {\it normal metal} (or lower superconducting gap) RF feeds. The high resistance of the device makes RF matching straightforward to achieve, and it can be read out using a wideband cryogenic HEMT amplifier. Klapwijk and Semenov~\cite{klapwijk2017engineering} provide an excellent description of the physics of operation with comprehensive references to performance. In this case, however, electron out-diffusion {\it increases} the effective electron-phonon interaction {\it volume}, increasing the achievable NEP. This intrinsic characteristic is at-odds with the ultra-low-noise requirement of the homodyne scheme being considered here.

\subsubsection{Low Temperature Hot Electron Bolometers}\label{sec:Low_temp_HEB} 

A hot electron bolometer operating at low temperature (nano-HEB) first seems to have been introduced by Karasik {\it et al.}\cite{karasik2011nanobolometers}, although the concept and operation can be viewed as an extrapolation to somewhat  higher operating temperature (whilst assuming extrapolation of behaviour to smaller TES lengths) of the electron-phonon decoupled TES first described by Cabrera {\it et al.}~\cite{cabrera1998detection}. Karasik {\it et al.} makes the same clear distinction between nano-HEB and diffusion-cooled HEBs as we do here. 

The nano-HEB relies on electron-phonon decoupling in a metal (N) connected to superconducting leads (S). Andreev reflection confines hot electrons within N, minimizing the interaction volume, and hence $G_b$ and the NEP. Proximity effects in N due to the S leads are often ignored in the description although expected in practice. The nano-HEB was envisaged to operate at  relatively high temperatures $T_c\sim 300\,{\rm mK}$ requiring small metal volumes $\lesssim 1\,{\rm \mu m^3}$, in order to achieve the best performance. By operating at a very low bath temperature, as demonstrated by Cabrera, the volume requirement is significantly relaxed and the usable dimensions much-increased. In what follows, we will describe an electron-phonon decoupled TES as a {\it nano-HEB} despite this background. 

\subsubsection{ Nano-HEB comments}\label{sec:comments_HEB}

The concept of the  nano-HEB  relies on Andreev reflection of quasiparticles at the S-N interface. Andreev reflection occurs if the S-N interface does not cause additional scattering: other than the reversal of the momentum of the in-coming sub-gap quasiparticle.  At the microscopic level, this seems to place restrictions on material and processing properties:

\begin{enumerate}

\item The S-N interface must not cause additional scattering.

\item There must be no out-diffusion of electrons from N, which would increase the effective interaction volume hence increase the  NEP.

\item The interface needs to be distinct at the scale size of the Fermi wavelength, which may be achievable with some S-N combinations. 

\item From a metallurgical viewpoint, particular requirements are placed on the binary phase-diagram of the constituent S-N films: no inter-grain diffusion and no intermetallics can be formed. Experience suggests that thin-film binary-metal combinations may still show additional unexpected effects beyond those seen in bulk phase-diagrams, perhaps arising from idiosyncrasies of the fabrication process.

\end{enumerate}
\noindent

These considerations suggest that extrapolations of nano-HEB performance from measurements on performance in relatively long structures to the short structures required for optimal performance, may not be reliable. Not-with-standing these concerns it is interesting to explore the potential of using electron-phonon decoupling for our application particularly if a low $T_c$  nano-HEB can be used.

In the original concept of the nano-HEB, operating at relatively high temperatures $T_c\sim 300\,{\rm mK}$  and for lowest NEP, the volume must be minimised so that, even for thin films (thickness $d\sim25\,{\rm nm}$) and widths $w\sim200\,{\rm nm}$, lateral dimensions of order $L\sim 1 \,{\rm \mu m}$ are required. The superconducting proximity effect from the leads would be expected to be significant, even for metals or alloys with short electronic mean-free-paths, and hence coherence lengths, giving much-reduced values for $\alpha$, compared to a TES \cite{harwin2023magnetic}. Later we will argue that this observation is much less of a concern for a nano-HEB designed to operate at $T_c\lesssim 100\,{\rm mK}$.

\subsection{{Josephson S-N-S} Detectors}\label{sec:SNS_detectors}

In this section, we overview the suitability of devices that rely on superconductor-normal metal-superconductor (SNS) proximity junctions. 

Govenius {\it et al.}\cite{govenius2014microwave, govenius2016detection} have described a threshold detector based on a series array of submicron-scale S-N-S Josephson proximity junctions. The sensitive element comprises a $30\,{\rm nm}$ thick AuPd absorber having dimensions that give a normal-state resistance  close to $36\,{\rm \Omega}$. The readout mechanism relies on the temperature dependence of the Josephson inductance in the series array of weak-links. In the mode of operation used,  devices were operated as threshold detectors (at $8.4\,{\rm GHz}$), rather than as bolometers, so that measurements or estimates of saturation powers are not available. 

Kokkoniemi {\it et al.}\cite{kokkoniemi2019nanobolometer} report measurements on the same, or at least, very similar, device. The readout is a resonant circuit operating at $\sim 8.4\,{\rm{GHz}}$. This would be of concern for axion homodyne detection as the readout tone is of the order of the signal frequencies of interest,  although it may be possible to shift the readout frequency with a redesign of circuit capacitances. Readout with a Josephson parametric amplifier, having an NEP of $6\times 10^{-20}\,{\rm{W/\sqrt{Hz}}}$ with a time constant $\tau\sim 30 \,{\rm\mu s}$ (BW $\sim 5.3\,{\rm kHz}$), was demonstrated. Power detection relies on pump-probe heating of the N-metal electron reservoir, but there are no values reported for saturation power. This means that estimates of performance for homodyne detection are not straightforward. However for scale, reported pump powers were of order $-126.5\,{\rm dBm}$ ($0.22\,{\rm fW}$) and so we take this to be an upper limit on the signal power at the frequency of interest, and assume a  power margin $\sim 0.5$ for the FoM estimate in Tab.~\ref{tab:FoM}. We could not rule-out this class of detector, but note some significant engineering effort might be needed to optimise the readout for the homodyne scheme.

\subsubsection{{Josephson SGS} Detectors}\label{sec:SGS_detectors}

The more recent work but related work of Kokkoniemi {\it et al.}\cite{kokkoniemi2020bolometer} using superconductor-graphene-superconductor (SGS) junctions extends this work. There an NEP $3\times 10^{-20}\,{\rm{W/\sqrt{Hz}}}$ with a time constant $\tau\sim 500 \,{\rm ns}$ (BW $\sim 310\,{\rm kHz}$) was reported. Saturation powers, inferred from Fig.~2(a) of the article were of order $P_{sat}\sim 0.2\,{\rm fW}$. 

\subsection{Thermal Kinetic Inductance Detectors}\label{sec:TKIDs}

It seems that Sauvageau and McDonald first suggested using the change in the kinetic inductance with temperature of the superconducting pairs in a thin film or wire as a mechanism for bolometric detection\cite{sauvageau1989superconducting}.
The idea has been revived more recently as the thermal kinetic inductance detector (TKID) \cite{wandui2020thermal}.
Recent TKID results describe the thermal characterisation of a TKID in a TES-like thermal geometry with power applied to a termination resistor and using SiN to give the required thermal isolation \cite{agrawal2021strong}. The TKID was designed to operate in a resonant circuit with a readout tone of order $318\,{\rm MHz}$, which may be sufficiently-below the signal frequencies of interest here.  Operating parameters were measured as a function of readout power (or DC bias power). $G_b$ and other thermal parameters were also measured. We have used these optimal parameters as inputs, with the {\it measured} NEP shown in Fig. 10 of the article, so as to calculate a FoM using Eq.~\ref{eqn:FoM}. Results are shown in column 4 of Tab.~\ref{tab:FoM}. 

We cannot rule out further optimization of this type of detector, as the performance is already impressive given the operating temperature $T_c=330\,{\rm mK}$. The potential hazard if trying to use a TKID at very low temperature must be electron-phonon decoupling. Quasiparticle lifetimes in TKIDs are expected to increase exponentially as the operating temperature is reduced so that the TKID becomes increasingly isolated from its SiN substrate and RF power input. Even so, operating a TKID at lower $T_c$ may be possible by using a lower-$T_c$  bilayer. A further concern is the possible increased contribution of two-level-system (TLS) noise, as described in Eq.~[16] of Ref.~\cite{wandui2020thermal}, as the operating temperature is reduced.

\subsection{Asymmetric STJ Detectors}\label{sec:S1IS2_detectors}

Superconducting tunnel junctions (STJs) have been used as energy-sensitive photon detectors for some considerable time.
They are formed from thin film superconducting films ($S$) separated by a very thin insulating barrier $I$. In the usual configuration the $S$ films are made of identical superconductors. Operated in a low magnetic field with suppressed Josephson supercurrent,  changes in quasiparticle densities in either S film can be detected from  time-resolved measurement of the tunnelling current and hence photon spectroscopy. More recently there has been renewed interest in asymmetric superconductor-insulator-superconductor ($S_1 I S_2$) junctions\cite{paolucci2023highly}. In this case, the bias circuit provides electrothermal feedback analogous to the electron cooling first described by Nahum~{\it et al.}\cite{nahum1994electronic} for S-N junctions and Manninen~{\it et al.}\cite{manninen1999cooling} for quasiparticles in $S_1-I-S_2$ junctions. The STJ described by Paolucci {\it et al.}\cite{paolucci2023thermoelectric} was formed from Al ($S_1$) and a superconductor/normal-metal Al/Cu bilayer ($S_2$) operated at $30\,{\rm mK}$. The superconducting energy gap of $S_2$ of $\Delta_{s2}= 80\,{\rm \mu eV}$ determines the detector threshold of $f = 19.3\,{\rm GHz}$. Operation at $f = 7.2\,{\rm GHz}$ with $T_b=10\, {\rm mK}$ is suggested, but estimates of NEP are not given.

\subsubsection{SIS Detectors}\label{sec:STJ}

SIS detectors are thin-film devices  that rely on the change in the Josephson supercurrent as the temperature of either film is increased. We have discounted SIS detectors (see e.g. Ref.~\cite{pankratov2022approaching}) as they operate as threshold detectors. Also, we do not consider SIS mixers as being suitable for homodyne detection at microwave frequencies, because the superconducting transition must be sharp on the energy of the local oscillator and signal being mixed, which implies that extremely sharp transitions are needed for microwave frequencies. Also the need for a low-noise IF amplifier increases the problems further. For these reasons, SIS mixer are never used at frequencies much below 100 GHz.

\section{Thermal Parameters for Modelling}\label{sec:Thermal_inputs}

\subsection{Thermal Isolation}\label{Thermal_isolation}

Over the years, a number of different configurations have been demonstrated to achieve the thermal isolation needed by TESs and HEBs. The conductance to the heat bath  $G_b$ can be engineered using either diffusive phonon transport, ballistic phonon transport, or electron-phonon decoupling.  

\subsubsection{Diffusive phonon isolation}\label{sec:Diffusive_phonon_isolation}

The thermally isolating support structures typically used for ultra-low-noise TESs are based on thin ($d\sim 200\,{\rm nm}$) silicon nitride (SiN) or silicon legs, with lengths $L\sim 100's \,{\rm \mu m}$ and widths $w\sim 1-10\,{\rm \mu m}$.  Our own measurements on thin, long SiN structures at temperatures $90<T_c<150\,{\rm mK}$ suggest that 
$K = \kappa (w d/L)$ where $\kappa\sim 1.1\times10^{-4}\,{\rm W/m\,K^n}$ and $n\sim 1.2$, for  widths $w=1-1.4 \,{\rm \mu m}$ and lengths $L>500\,{\rm \mu m}$. Leivo and Pekola\cite{leivo1998thermal, pekola2021colloquium} measured a different behaviour to this and to a slightly lower temperature ($70\,{\rm mK}$). Their results were parameterized in terms of $G=k_{LP}\left(A/L\right)T^m$ where $m=n-1$ (in the present notation), and $A=wd$ is the cross-sectional area.  In our nomenclature $k_{LP}=n\kappa$. Comparative values are given in Table~\ref{tab:Conductance}. In the following modelling of TES performance we used the  Leivo-Pekola value measured for full membranes (Geometry I in Table~\ref{tab:Conductance}). 

It is often found that ultra-low noise  bolometers that rely on diffusive isolation show quite variable bath-conductances even between nominally identical devices fabricated on the same wafer in close physical proximity. This may arise from the nature of the scattering mechanisms and localisation of the microwave phonons that determine the heat flow in constricted bridges at low temperatures, where phonon wavelengths become comparable to the dimensions of the structure and/or its defects.\cite{withington2011low}
\begin{table}
\centering
\begin{tabular}{lccccc}
\toprule
\textbf{} & $w\times L$ & $k_{\textrm{LP}}$ & $m$ & $\kappa$ &$n$\\
  & ${\rm \mu m}$ &${\rm W/m\,K^m}$&  & ${\rm W/m\,K^n}$ & \\
\midrule
Geom. I\cite{pekola2021colloquium} & Full membrane & $1.45\times10^{-2}$ & 1.98 &$4.87\times10^{-3}$ & 2.98\\
Geom. II& $25\times 100$   &  $1.58\times10^{-3}$ & 1.54 & $6.22\times10^{-4}$ & 2.54\\
Geom. III&$4\times 100$  & $5.70\times10^{-4}$ & 1.37 & $2.41\times10^{-4}$ & 2.37\\
Our measurements &  $1.4\times>500$ & &  & $1.1\times10^{-4}$ & 1.2\\
\bottomrule
\end{tabular}
\caption{\label{tab:Conductance}
Comparative values for diffusive thermal conductance parameters used in this study.  
}
\end{table}
\begin{figure}
\centering
\includegraphics[width=8.6cm]{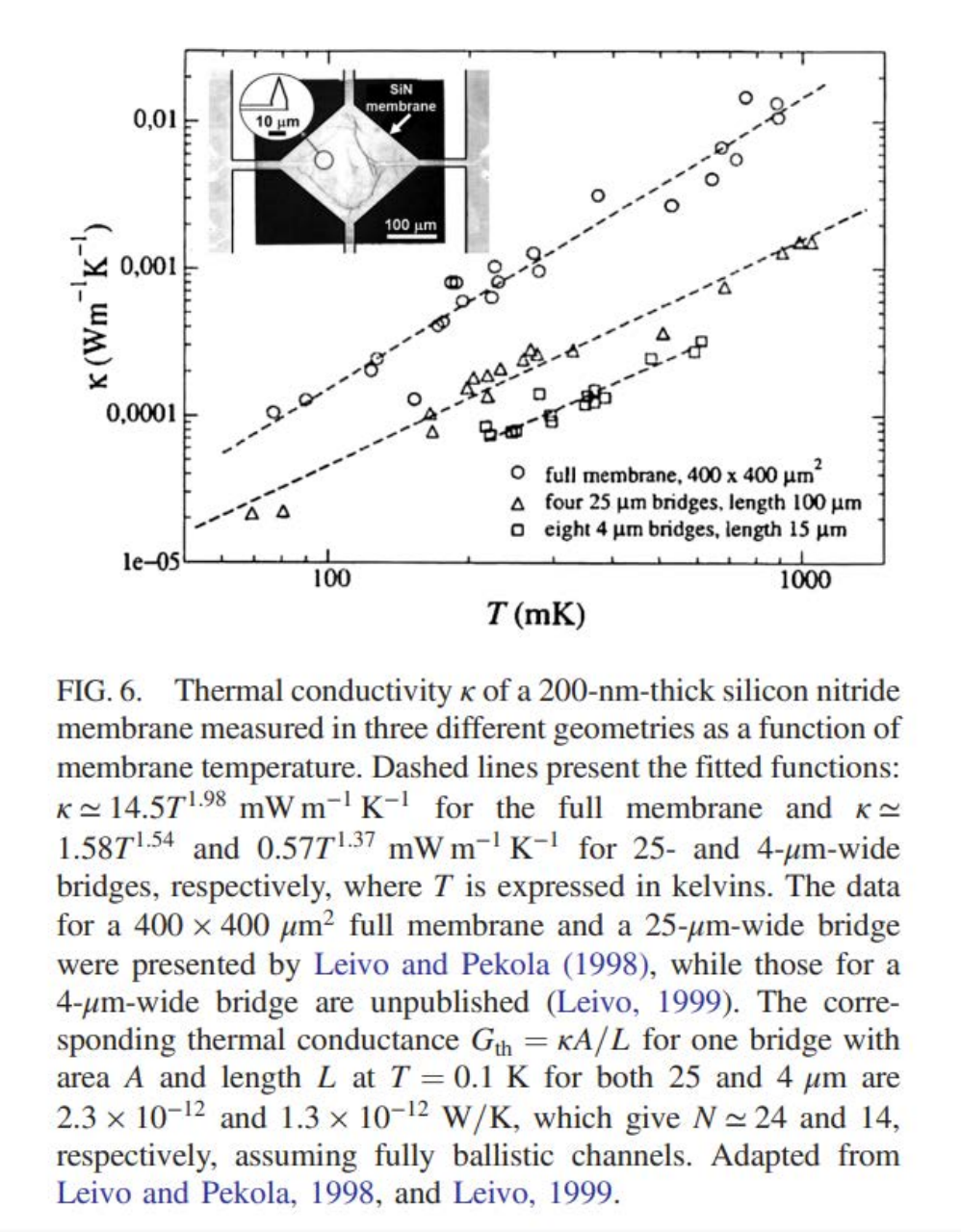}
\caption{\label{fig:Leivo_Pekola} Results taken from Leivo and Pekola\cite{leivo1998thermal}, 
Pekola {\it et al.}\cite{pekola2021colloquium} showing the thermal conductivity of SiN membranes in 3 geometries to $70\,{\rm mK}$}
\end{figure}

\subsubsection{Ballistic phonon isolation}\label{sec:Ballistic_isolation}

For short narrow dielectric structures, heat propagation at low temperature becomes ballistic and is limited by the four lowest acoustic massless modes of the thermal link (i.e. those with zero energy threshold \cite{pekola2021colloquium,schwab2000measurement}. These four lowest modes arise from one dilational, one torsional and two flexural degrees of freedom. For ideal coupling between the ballistic channel and the heat sinks (the island supporting the TES and the heat bath of the Si substrate) the power flow per mode  $P_{q}$ is given by
\begin{equation}\label{eqn:P_q}
	P_{q}= \frac {\pi^2k_B^2} {6h}\left( T_c^2 - T_b^2 \right),
\end{equation}
and the conductance per mode is 
\begin{equation}\label{eqn:G_q}
	G_{q}=\frac {\pi^2k_B^2T_c} {3h}.
\end{equation}

In the low-temperature limit the ballistic conductance per link is $G=4G_q$. Ballistic phonon isolation has been measured by Schwab\cite{schwab2000measurement} and  by Osman\cite{osman2014transition} in the context of TES thermal isolation. Williams\cite{williams2018superconducting} showed enhanced isolation in short thermal links using micromachined phonon interferometers. 

\begin{figure}
\centering
\includegraphics[width=8.6cm]{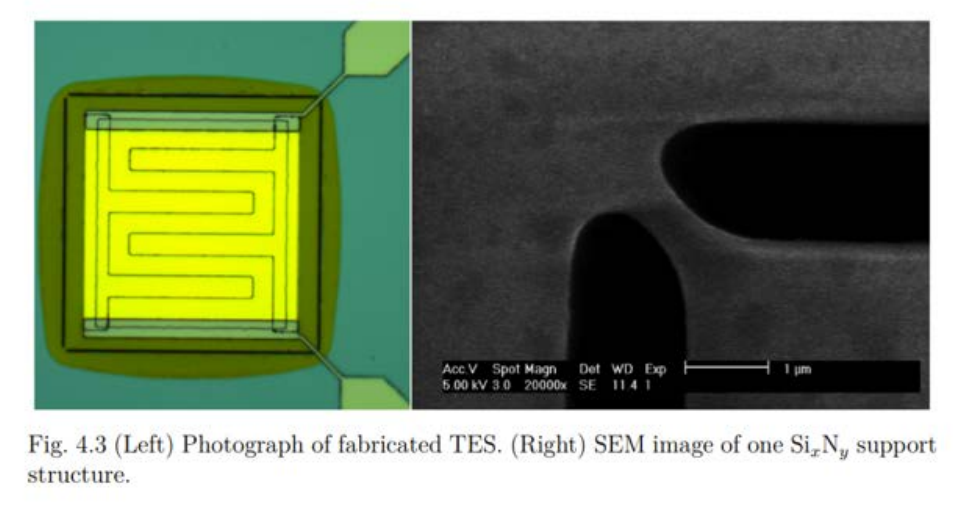}
\caption{\label{fig:Osman_ballistic} Images of a ballistic-isolated TES from Osman {\it et al.}\cite{osman2014transition}}
\end{figure}

The key point is that ballistic phonon isolation yields highly reproducible thermal characteristics. The exponent in the power-flow is $n=2$ for ballistic propagation, and only four modes propagate at low temperatures. These represent almost ideal characteristics for a moderate NEP device.

\subsubsection{Electron-phonon decoupling}\label{sec:e_ph}

In thin metal films (thickness $d\lesssim 200 \,{\rm nm}$) at low temperatures $T<1\,{\rm K}$, electrons and phonons  become increasingly decoupled \cite{ pekola2021colloquium, wellstood1994hot, roukes1985hot}. The power flow $P_{e-ph}$ between the electron and phonon sub-systems is usually described using the power law $P_{e-ph} = \Sigma V\left(T_e^n - T_{ph}^n\right)$, where $\Sigma$ is a material-dependent parameter, $V$ is the  volume of the metal, and $T_{e,ph}$ characterise the electron and phonon temperatures respectively. The electron-phonon conductance $G_{e-ph}=dP_{e-ph}/dT_e$ is then given by
\begin{equation}\label{eqn:G_e_ph}
	G_{e-ph}=n\Sigma V T_e^{n-1}.
\end{equation}
For clean metals with large electron mean-free-paths, $\lambda_e\gg d$, $n\simeq 5$, although this appears to be modified by disorder.

\subsubsection{Thermal boundary resistance}\label{Kapitza}

Thermal mismatch  across a boundary between two  insulating materials means that the temperatures of the phonon gases on either side of the boundary $T_{ph1,ph2}$ are different when power flows. This behaviour is usually described by defining a thermal conductance between the films: often termed the Kapitza conductance $G_{K}$. The power flow across a boundary is usually parameterized by
\begin{equation}\label{eqn:P_k}
	P_K=\frac{1} {4} \sigma_K A_I (T_{ph1}^4-T_{ph2}^4),
\end{equation}
where $A_I$ is the area of the interface, and then the associated conductance is given by 
\begin{equation}\label{eqn:G_Kap}
	G_{K}=\sigma_{K} A_I T^3.
\end{equation}
Although $\sigma_K$ varies between materials, a value of  $\sigma_K=500\,{\rm Wm^{-2}K^{-4}}$ is generally sufficient for first-order numerical calculations \cite{roukes1985hot,elo2017thermal}.

\subsection{Thermal conductance in metals}\label{G_WF}

The  thermal conductance of metal films with  length $L$ and cross-sectional area $A$ can be calculated using  the Weidemann-Franz Law so that 
\begin{equation}\label{eqn:G_WF}
	G_{WF}=\frac{A L_0 T} {L\rho } ,
\end{equation}
where $L_0=2.44\times10^{-8}\,{\rm V^2 K^{-2}}$ is the Lorentz number and $\rho$ is the film resistivity.

\subsection{Heat capacities}\label{sec:Heat_capacities}

\subsubsection{Metals}\label{sec:C_metals}

At low temperature, the main contribution to the specific heat capacity of a metal arises from the electron system. The electron specific heat capacity per unit volume is
\begin{equation}\label{eqn:C_e}
	c_{e}=\gamma  T_e,
\end{equation}
where $T_e$ is the electron temperature. Values of  $\gamma$ are tabulated: for example see Ref~\cite{Gladstone}). For a superconductor close to its transition temperature $T_c$, as described by BCS, the specific heat is enhanced by a factor of up to 2.43 compared with the normal-metal value due to the temperature-dependence of the energy gap in the electronic excitation spectrum. 

\subsubsection{Crystalline dielectrics}\label{sec:C_dielectric}

The specific heat capacities of crystalline dielectrics at low temperatures can be calculated using the Debye model: 
\begin{equation}\label{eqn:C_ph}
	c_{ph}=1944 \left(\frac{T}{\Theta_D}\right)^3 \, {\rm J/[mol]\,K},
\end{equation}
Again, Ref.~\cite{Gladstone} provides a useful table for the Debye temperature $\Theta_D$. 

\subsubsection{Amorphous dielectrics}\label{sec:C_amorphous}

The specific heat capacity of amorphous dielectrics at low temperatures is largely  determined by two-level-systems. We have measured specific heat capacities in SiN comparable to those of normal metals with a linear dependence of $c_{\rm SiN}$ on  temperature \cite{rostem2008thermal}. This process is particularly important when carrying out thermal modelling of TESs on SiN.  

\subsection{TES and nano-HEB response times}\label{TES_response_time}

\begin{table}
\centering
\begin{tabular}{ccccccc}
\toprule
\textbf{Material} & $T_c$       & $\Sigma$                   & $n$   &$\gamma $               & $\tau_{e-ph}(T_c)$   & \textbf{Ref} \\
                          & $\rm{mK}$ &$\rm{W/{m^3K^n}}$&        &$\rm{J/{m^3K^2}}$ & $\rm{\mu s}$&                    \\
\midrule
        W              &  95             &  $4.67\times 10^7$   &  5     &  136					& 680        &  \cite{cabrera1998detection}     \\
        Ti              &  359           &  $6.22\times 10^8$   &  5     &  310					& 2.2       &  \cite{fukuda2007titanium}     \\
        Ti              & 510          &  $1.30\times 10^9$   &  5     &  310					& 0.4         & \cite{manninen1999cooling}     \\
        Ti              &  351          &  $2.05\times 10^8$   &  4     &  310					& 3.1        &  \cite{taralli2007development}  \\
        Hf              &  300         &  $1.65\times 10^6$   &  6     &  160					& 2000        &  \cite{karasik2011nanobolometers}   \\ 
        Ir               &  114         &  $ 5.92\times 10^7$   &  5     &  380					& 870        &    \cite{karasik2011nanobolometers}       \\ 
       TiAu           &  96         &  $ 1.11\times 10^{10}$   &  5     &  85.9					&3.9        &      \cite{taralli2007development}   \\
       Au              & 100        &  $ 2.4\times 10^9$   & 5        & 73.4                           & 6.1        &     \cite{echternach1992electron}      \\
       AuPd           &  100        &  $ 3.0\times 10^9$   &  5     &  579					&48.3        &    \cite{govenius2014microwave}     \\
       AuCu          &  100        &  $ 2.4\times 10^9$   &  5     &  85.4					&7.1        &      \cite{wellstood1989hot}              \\   
\bottomrule
\end{tabular}
\caption{\label{tab:nano-HEB inputs} Values adopted in this paper, characteristic of the performances of nano-HEBs, following Giazotto\cite{giazotto2006opportunities} and Karasik\cite{karasik2011nanobolometers}. Updated  material parameters are included where available. Note $\tau_{e-ph}$ is calculated at the reported $T_c$. For normal metals we assume a reference value of $T_c=100\,{\rm mK}$. 
}
\end{table}

A TES or nano-HEB comprises  a low-$T_c$ superconductor ($S^{\prime}$)  with electrical contacts formed from a higher $T_c$ superconductor ($S$).  The TES is voltage-biassed, and read out using a SQUID ultra-low-noise current to voltage transformer. The operation of the TES relies on the rapid change of resistance of $S^{\prime}$ close to its transition temperature. The sharpness of the change of its resistance $R$ with temperature $T$ is characterized by the dimensionless parameter $\alpha=(T/R) dR/dT$. The change in resistance occurs over a narrow temperature range around $T_c$, and so $\alpha$ can be large $\sim 100-200$. Voltage bias creates ETF, which amongst other effects, means that the effective response time of a TES or HEB $\tau_{eff}$ is reduced from its open-loop value  $\tau_0=C_{TES}/G_b$ such that $\tau_{eff}\simeq \tau_0/(1+\alpha/n)$ \cite{Irwin2005}. The underlying approximations assume that the change in $R$ with current $I$, $\beta=(I/R) dR/dI$ is small: for suitable superconductors $\beta \sim 0.1 \alpha$ . Large values of $\beta$ slow down the response, leading to narrow readout bandwidths. For later thermal modelling we assume $\beta\sim 0$.

Table~\ref{tab:nano-HEB inputs} combines and updates where available Table~1 of Ref.~\cite{giazotto2006opportunities} and 
Table~1 of Ref.~\cite{karasik2011nanobolometers} giving the material parameters  $\Sigma$, $n$ and $\gamma$  used for estimating the suitability of using ultra-low temperature TESs and nano-HEBs for microwave homodyne detection.
Note that Ref.~\cite{karasik2011nanobolometers} defines $n$ such that $\tau_{e-ph}\sim T^{-n}$. We use $n$ as the exponent in the power flow as in Eq.~\ref{eqn:P_bath} and as in Refs.~\cite{giazotto2006opportunities,wellstood1989hot}
(this approach makes discussion and comparison  of different power-flow mechanisms simpler) so that $n_{\rm here}=n_{\rm Karasik}+2$.  Values of $\tau_0$ are calculated at the reported $T_c$ for each material, other than for the normal metals where we have assumed a nominal $T_c=100\,{\rm mK}$. Values for normal metals (and alloys) are included here for later TES performance estimates as combinations in bi- or multi-layers. We have excluded superconductors having $T_c\ge 500\,{\rm mK}$. 

\subsection{TES and nano-HEB readout and normal-state resistance limits}\label{sec:Rn_TES_max}

A TES or HEB is  a low-resistance device having a normal-state resistance of $R_n\sim0.1-0.5\,{\rm \Omega}$. Read out with a SQUID places a requirement on the normal-state operating resistance of the TES. For a TES or nano-HEB for homodyne detection with probe power, we assume $R_0\sim R_n/2$. The broadband SQUID readout current noise is typically of order $i_{n,sq}\sim 4\,{\rm pA/\sqrt{Hz}}$. Irrespective of the thermal decoupling mechanism used, we need to ensure that the current  noise due to the thermal conductance to the heat-bath $i_{n,G_b}$ is greater than $i_{SQ}$, by a factor of about 3. This means that the thermal decoupling determines the achievable NEP when the other noise sources are added in quadrature. The TES is voltage biassed at $V_0$ so that 
\begin{equation}\label{eqn:In_Gb_1}
	i_{n,G_b}  =   \frac  {\sqrt {4k_bT_c^2G_b}}  {V_0 }.
\end{equation}
Recognising that $V_0^2/ R_0 = P_{b}$, and assuming a TES operating  at $R_0=R_n/2$, Eq.~\ref{eqn:In_Gb_1} can be simplified  to give
\begin{equation}\label{eqn:Rn_max}
	R_{n,\max} \le  \frac {8k_b n T_c} {9 i_{n,sq}^2 \phi(T_c,T_b) },
\end{equation}
which places an upper limit on the normal-state resistance of any TES read-out using conventional DC SQUIDs.  For a TES with ballistic phonon isolation ($n=2$) and $T_c=30\,{\rm mK}$, we find $R_{n,\max} \sim 60 \, {\rm m\Omega}$.  For a nano-HEB with electron-phonon   ($n=5$) and $T_c=30\,{\rm mK}$, we find $R_{n,\max} \sim 120 \, {\rm m\Omega}$.
 
\subsection{Illustrative design of a TES for Homodyne Detection}\label{TES_performance}

Given the technological maturity, it is relatively straightforward to imagine how a single-chip integrated homodyne detector could be fabricated using two TESs. Each TES would be thermally isolated by using a patterned dielectric film. Power would be coupled into a separate load resistor that terminates a superconducting thin-film microstrip: a technology that we have refined to high TRL and reproduced many times. The termination resistor would be designed to be impedance-matched to the microstrip impedance so that $R_{res}\sim 25\,{\rm \Omega}$ for microstrip coupling or $R_{res}\sim 25\,{\rm \Omega}$ for CPW. For thermal isolation we would choose ballistic phonon isolation so that the realised conductance to the heat bath is well-defined at low temperatures. We assume that four thermal channels connect to the heat bath in each support leg at all temperatures: this assumption becomes increasingly inaccurate at higher operating temperatures as additional modes cut-on,\cite{osman2014transition}, but is not important in the case of the low-temperature application considered here. We assume that a low temperature heat bath at $T_b=15\,{\rm mK}$ is available, implying a target $T_c\simeq 30\,{\rm mK}$. A superconducting bilayer or multi-layer with appropriate  $T_c$ then needs be designed for the purpose. We show more detail of a possible geometry and results of detailed thermal modelling  in Sec.~\ref{sec:TES_for_homodyne}. For a TES fabricated on SiN we have explored two geometries: a detector where the TES is laterally separated by the power-absorbing resistor ({\em distributed TES}) and a geometry where the power absorbing resistor is fabricated on top of the TES using, undesirably, additional dielectric isolation  ({\em stacked TES}). The latter has interesting possibilities because the coupling between the power-absorbing element and the TES should be determined by the Kapitza conductance across interfaces rather than the slower diffusive conductance of SiN. 

A particularly interesting possibility is a TES with ballistic phonon isolation fabricated using short crystalline Si legs. Such a device could in principle be fabricated from a silicon-on-insulator (SoI) wafer with a suitably thin crystalline Si device layer $d\sim 200\,{\rm nm}$. We expect that this scheme would significantly reduce the heat capacity and improve the the thermal conductance, without increasing the noise, which would lead to large readout bandwidths. Provided the thermal isolation has the reduced dimensions discussed previously, the thermal conductance remains ballistic and well-defined, without the complicating effects of localised phonon transport. There may also be significant advantages associated with the expected reduction in the contribution of TLSs to the device's heat capacity and improved thermalization across the TES island. 

\subsection{Illustrative design of a nano-HEB for Homodyne Detection}\label{HEP_performance}


\begin{table}
\centering
\begin{tabular}{cccccc}
\toprule
\textbf{Parameter} &                                  &  \textbf{Au}   &  \textbf{AuCu} &          \textbf{AuPd}       & \textbf{Ti}  \\
\midrule
        $\Sigma$      & $\rm {W/m^3 K^n}$  &  $ 2.4\times 10^9$  & $ 2.4\times 10^9$    &  $ 2.4\times 10^9$ & $ 1.3\times 10^9$ \\
        $\gamma$      & $\rm{J/{m^3K^2}}$  &  $ 73.4$  & $ 85.4$    &  $579$ & $310$ \\
        $n$              &                                    &  5                            &  5     &  5     				& 5              \\
        $T_c$          &     ${\rm mK}  $           &  30                         &  30      &  30				      & 30             \\
        $V_{\max} $  &  ${\rm \mu m^3}  $     &  46.6          & 46.6                       &  46.6					         &             \\
         $\rho$      &   ${\rm n\Omega \, m} $ &  $13.3 $&  $72.0 $   &  234.0     &  200.0     \\
        $d$              &  ${\rm \mu m}  $       &  0.1  &  0.1    & 0.1					&         \\ 
       $w$               &  ${\rm \mu m}  $       & 20  &  50     &  80					&         \\ 
       $L$               &    ${\rm \mu m}  $      &   18   &  8     &  4					&         \\

       $R_n$           &   ${\rm m\Omega} $      &  120  &  115     & 120					&            \\
       $\xi_n(T_c)$           &  ${\rm \mu m}  $        &  1.17   &  0.46     &  0.23			&        0.08    \\
        $\tau_{e-ph}(T_c)$           &  ${\rm \mu s}  $        &  230  &  260     &  1800		&       30   \\
      $P_{sat}$          & ${\rm fW}  $        &  2.63  &   2.63      &  2.63 					&             \\   
\bottomrule
\end{tabular}
\caption{\label{tab:HEB concept}
Preliminary design parameters for an electron-phonon decoupled nano-HEB needed to achieve ${\rm NEP}=150  \,{\rm zW/\sqrt{Hz}}$ operating at $T_c=30\,{\rm mK}$ based on N-S multilayers.
}
\end{table}

In this section, we assume that nano-HEBs can be used as the power absorbing elements of homodyne detectors; an assumption that will be considered further in  Sec.~\ref{sec:Impedance_matched _HEB}. Table~\ref{tab:HEB concept} shows preliminary design estimates for an electron-phonon decoupled nano-HEB designed to achieve  ${\rm NEP}=150\,{\rm zW/\sqrt{Hz}}$. As illustrative parameters, we use $T_c=30\,{\rm mK}$ and $n=5$ for all materials.  Using Eq.~\ref{eqn:Rn_max} we need $R_{n,\max}\sim 120\,{\rm m\Omega}$ at $30\,{\rm mK}$, which is satisfied. In all cases, the chosen $L/ \xi_n (T_c)\sim 17$ in order to minimise the proximity effects of the leads; where we expect the proximity effects to scale as $\exp(-L/\xi_n)$. We have included partial entries for Ti, as this may be a suitable choice for the S-material  if
engineered to have $T_c=30\,{\rm mK}$ as part of a S/N multilayer. The relatively short electron-phonon scattering time in Ti and its relatively high resistivity (other than compared to AuPd) means that the N-layer component would be expected to dominate both the effective response time and resistance. A refinement of this restriction would need more detailed modelling. 

Table~\ref{tab:HEB concept2} shows preliminary design estimates for an electron-phonon decoupled TES designed to achieve 
${\rm NEP}=50\,{\rm zW/\sqrt{Hz}}$.
Again, we assume $T_c=30\,{\rm mK}$ and $n=5$ for all materials. 
For this lower NEP detector the ratio  $L/\xi_n (T_c)\sim 9 $ is still expected to give minimal proximity effects from the leads. 
\begin{table}
\centering
\begin{tabular}{cccccc}
\toprule
\textbf{Parameter} &                                  &  \textbf{Au}   &  \textbf{AuCu} &          \textbf{AuPd}       & \textbf{Ti}  \\
\midrule
        $\Sigma$      & $\rm {W/m^3 K^n}$  &  $ 2.4\times 10^9$  & $ 2.4\times 10^9$    &  $ 2.4\times 10^9$ & $ 1.3\times 10^9$ \\
         $\gamma$      & $\rm{J/{m^3K^2}}$  &  $ 73.4$  & $ 85.4$    &  $579$ & $310$ \\
        $n$              &                                    &  5                            &  5     &  5     				& 5              \\
        $T_c$          &     ${\rm mK}  $           &  30                         &  30      &  30				      & 30             \\
        $V_{\max} $  &  ${\rm \mu m^3}  $     &  5.18          &  5.18                        &   5.18 				         &             \\
         $\rho$      &   ${\rm n\Omega \, m} $ &  $13.3 $&  $72.0 $   &  234.0     &  200.0     \\
        $d$              &  ${\rm \mu m}  $       &  0.1  &  0.1    & 0.1					&         \\ 
       $w$               &  ${\rm \mu m}  $       & 10  &  25     &  40					&         \\ 
       $L$               &    ${\rm \mu m}  $      &   12   &  4     &  2					&         \\

       $R_n$           &   ${\rm m\Omega} $      &  130 &  115     &  		120		&            \\
       $\xi_n(T_c)$           &  ${\rm \mu m}  $        &  1.17   &  0.46     &  0.23			&        0.08    \\
        $\tau_{e-ph}(T_c)$           &  ${\rm \mu s}  $        &  230  &  260     &  1800		&       30   \\
      $P_{sat}$          & ${\rm fW}  $        &  0.29   &   0.29      &  0.29					&             \\   
\bottomrule
\end{tabular}
\caption{\label{tab:HEB concept2}
Preliminary design parameters for an nano-HEB to achieve ${\rm NEP}=50  \,{\rm zW/\sqrt{Hz}}$ operating at $T_c=30\,{\rm mK}$ based on N-S multilayers.
}
\end{table}
Note that the FoM described in Eqs.~\ref{eqn:FoM} or \ref{eqn:FoM2} is unchanged as the NEP is reduced. 
AuCu seems to be particularly attractive, given its low heat capacity,  hence short $\tau_{e-ph}$ time, and its relatively high resistivity  hence short coherence length. We have modelled a nano-HEB with a AuCu multilayer in Sec.~\ref{sec:HEB_performance}.

\section{Estimates of Figures of Merit for Candidate Detectors}\label{sec:FoM}

We have used Eq.~\ref{eqn:FoM2} to account for bolometric detectors, or Eq.~\ref{eqn:FoM} for the general case, to estimate the figures-of-merit of some of the potentially competing superconducting detector technologies. Results from the calculations are shown in Table~\ref{tab:FoM}, where the FoM was calculated for $\nu_p=10\,{\rm GHz}$ in all cases. 

\begin{table}
\centering
\begin{tabular}{lcccc}
\toprule
\textbf{Parameter} & \textbf{TES} & \textbf{HEB} & \textbf{SNS} &\textbf{TKID} \\
\midrule
 $P_{b}\,({\rm fW}$) & 6.4 & 1.97 & 0.22 &  3750 \\
 ${\rm NEP}\, ({\rm {zW/\sqrt{Hz}}}$) & 145 & 150 & 60 &  25000\\
 $T_c\,({\rm{mK}}) $& 30 & 30 & 25 & 330 \\
  $n$ & 2 & 5 & 5 & 3.2 \\
 $\gamma$ & 0.75 &0.75 &  & 0.75  \\
 $\kappa_{pump}$ & 0.5 & 0.5& 0.5 & 0.5 \\
 $\kappa_{rf}$ & 1 & 1 & 1 & 1 \\
  $\phi (T_b,T_c) $ & 0.75 & 0.97  & &  0.84 \\
\textbf{FoM}  & \textbf{2.00} & \textbf{1.03} & \textbf{0.40} & \textbf{0.08} \\
\bottomrule
\end{tabular}
\caption{\label{tab:FoM}
Estimates of comparative values for the FoM for representative detector technologies identified in this study.  
The FoM is calculated for a modulation frequency $\nu=10\,{\rm GHz}$ in all cases.
}
\end{table}
\begin{figure}
\centering
\includegraphics[width=8.6cm]{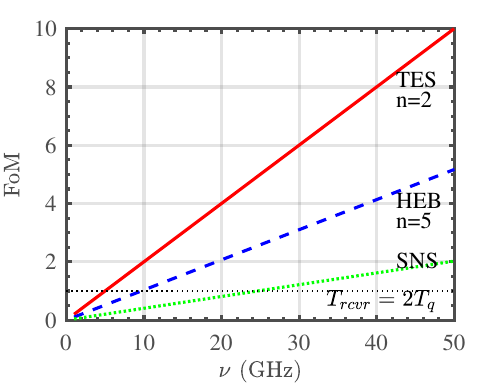}
\caption{\label{fig:FoM_v1} Plots of the calculated FoM for bolometers with different thermal isolation and for the SNS bolometer 
described in Ref.~\cite{kokkoniemi2019nanobolometer}.}
\end{figure}

Figure~\ref{fig:FoM_v1} shows plots of calculated FoMs of bolometers with different styles of thermal isolation and for the SNS bolometer  described in Ref.~\cite{kokkoniemi2019nanobolometer}. For the bolometers, we use Eq.~\ref{eqn:FoM2} while for the SNS we use reported values for the NEP as in Eq.~\ref{eqn:FoM}. For a TES with ballistic phonon thermal isolationoperating at $30\,{\rm mK}$, homodyne detection becomes quantum-limited at frequencies above $5.0\,{\rm GHz}$. For a TES with electron-phonon decoupling operated at $30\,{\rm mK}$ this crossover frequency is increased to $9.7\,{\rm GHz}$. Fr the SNS bolometer operating at $30\,{\rm mK}$ we estimate $49.3\,{\rm GHz}$. 

\goodbreak
\section{Thermal modelling of TESs for Homodyne Detection}\label{sec:TES_for_homodyne}

\subsection{Overview of device designs}\label{sec:tes_designs}

We consider the thermal performance of three different TES bolometer designs. These have been developed by modifying our existing TES technology to meet the requirements of very low-temperature homodyne operation. The numerical modelling tools have been developed over many years, and have been shown to be good indicators of actual experimental behaviour.

All three types share a similar basic configuration with a TES, RF-power absorbing resistor and phononic thermal isolation. The  designs differ in the layout of the TES and resistor, and membrane material used. In the first two types, the membrane is assumed to be low-stress amorphous SiN. They differ in the layout of the island: in the distributed design the resistor and TES are placed side-by-side, whereas in the stacked arrangement they are placed on top of each other on either side of an insulating SiO$_2$ spacer layer. The third design also uses a distributed layout, but assumes a crystalline Si membrane in place of SiN.

Wherever possible we fix the TES operating parameters in all simulations so that $R_n = 60\,{\rm m\Omega}$, with operating resistance $R_0=15\,{\rm m\Omega}$, and the TES operating current point $I_0$ satisfies $I_0^2R_0=\Sigma_{i=1\dots Nch} P_{b,i}$, where the sum is over the number of thermal channels to the heat bath; for example, $N_{ch}=5$ in the case of ballistic isolation with 5 thermal legs described above. Unless otherwise stated $T_c=30\,{\rm mK}$. The ETF parameters are taken to be  $\alpha=200$, $\beta=0.01$ in order to describe a TES with a sharp resistive-temperature transition and having little effect from critical current limitations.This simplifies the interpretation of the predicted power responsivity etc., and is realisable in practice: the TES itself should not place a limit on the readout bandwidth. With this choice of TES values, and a realistic inductance for the SQUID input inductance $L_{in}=80\,{\rm nH}$ regularly measured by us, the electrothermal dynamics of the TES are expected to be close-to critically damped.

We show results of the effects of direct power dissipation in the TES bilayer, which allows the thermal behaviour to be better understood, and in the effects of power dissipation in the RF load resistor, which is the intended mode of operation. The former confirms that our numerical modelling agrees with analytical descriptions in the appropriate limits, the latter indicates the predicted overall performance. Further details of the electrothermal modelling are given in Refs.~\cite{figueroa2006complex} and \cite{goldie2009thermal}.


\subsection{Distributed TES and separate load resistor on SiN}\label{TES_dist_load}

\begin{figure}
\centering
\includegraphics[width=7cm]{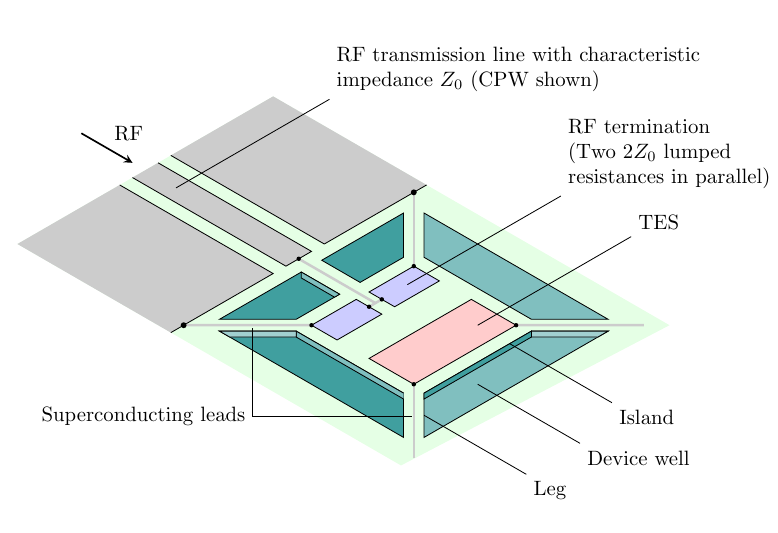}
\caption{\label{fig:TES_concept} Concept for a coplanar waveguide-coupled TES detector with balanced termination resistors on thin-film SiN (light green areas). The dark green areas highlight  regions that are removed by etching to provide thermal isolation. }
\end{figure}

Figure~\ref{fig:TES_concept} shows a conceptual design of a coplanar waveguide-coupled TES detector with balanced termination resistors on thin-film SiN (light green regions). The dark green areas highlight the etched regions providing thermal isolation. The TES and RF load resistor are fabricated together on a 200-250\,nm thick membrane suspended across a micro-machined window in the host Si chip. The membrane is patterned to form an island that is connected to the chip by five narrow support legs that ensure ballistic phonon transport. This arrangement is a straightforward modification of the TES processing route we have been using for many years to fabricate ultra-low-noise FIR and microstrip-coupled devices.

The island hosts the TES and RF termination resistor and together account for most of the heat capacity of the bolometer, with micromachined legs providing the thermal conductance to the chip, which plays the role of thermal bath. Our previous measurements show that this can be achieved for SiN by using legs that are $200\,{\rm nm}$ thick, 1.5~$\mu$m wide and $\sim1-4 \,\mu$m long \cite{osman2014transition}. RF power from the input transmission line is dissipated in the termination resistor, providing a thermal signal power to the island that is sensed by the TES bilayer.

\begin{figure}
\centering
\includegraphics[width=7cm]{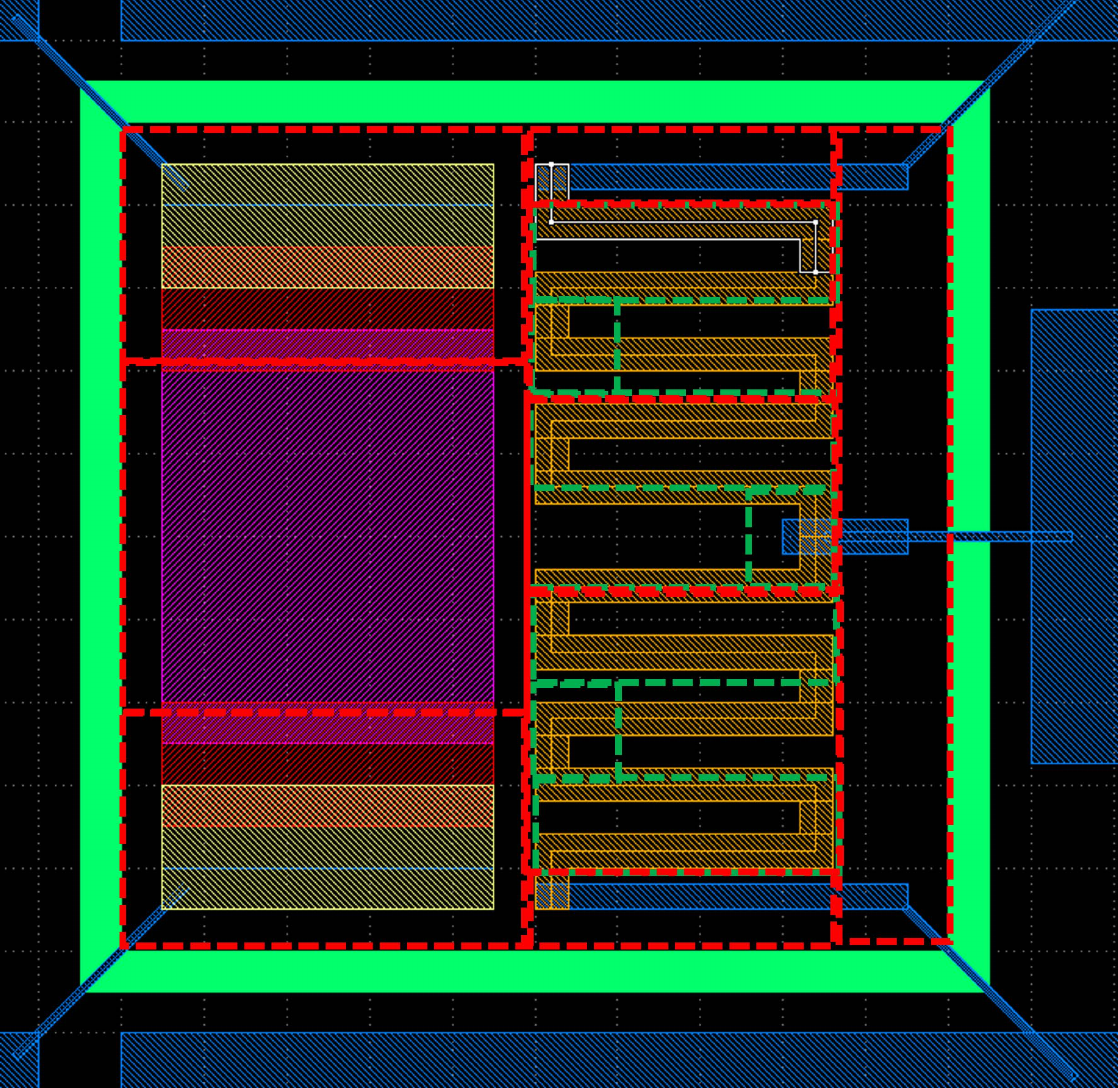}
\caption{\label{fig:Mask_v1} Plan view of a draft photolithography mask layout for a distributed TES
 as shown in Fig.~\ref{fig:TES_concept}.}
\end{figure}

\begin{figure}
\centering
\includegraphics[width=7cm]{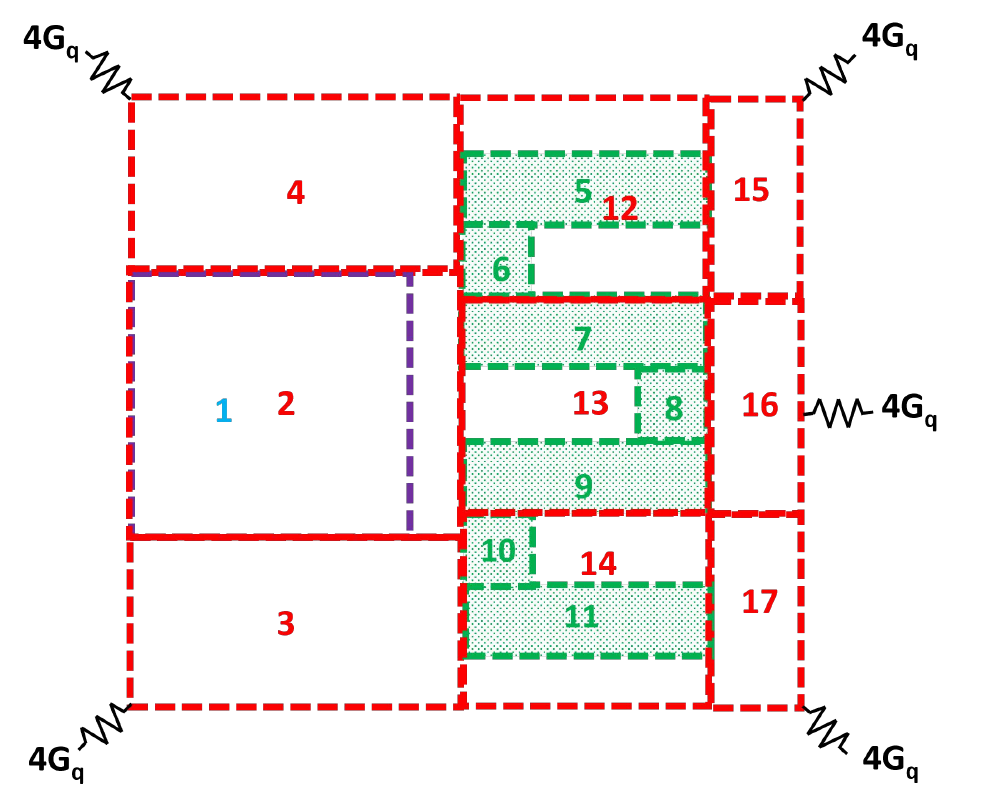}
\caption{\label{fig:Thermal_v6} Heat capacities used to approximate the TES design shown in Fig.~\ref{fig:Mask_v1}. Components in purple denote the TES itself, red phonons and green the resistor electrons.}
\end{figure}

\begin{figure}
\centering
\includegraphics[width=10cm]{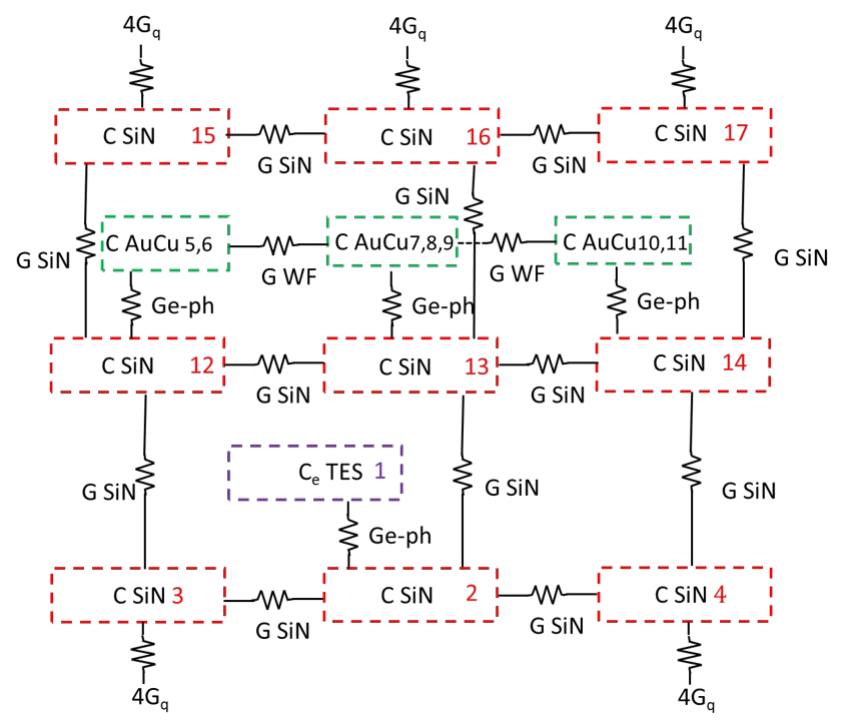}
\caption{\label{fig:Thermal_v6_model}  Schematic of a thermal model for a distributed TES. Components in purple denote the TES itself, red phonons an green the resistor electrons.}
\end{figure}

By way of example, Fig. \ref{fig:Mask_v1} shows a draft photolithography mask that would fabricate  a  distributed TES and termination resistors on SiN with ballistic phonon thermal isolation. Figures~\ref{fig:Thermal_v6}  and \ref{fig:Thermal_v6_model} show the thermal model used to create the design shown in Fig.~\ref{fig:Mask_v1}.

For the thermal modelling work described here, it is necessary to assume representative dimensions for the various elements. The dimensions of the island are determined by the requirement that it must be large enough to host the TES and resistor. We assume a TES having dimensions $w\times L = 40\times 40\,{\rm \mu m^2}$ ($w$ is the TES width, $L$ is the TES length) based on previous work, with a representative thickness $d_{TES}=250\,{\rm nm}$. The latter is consistent with the thickness needed for a bi- or trilayer comprising Au and Ti in order to achieve an operating temperature of $T_c=30\,{\rm mK}$. In the case of the RF resistor, it must terminate, on leaving the chip, a 50\,$\Omega$ feedline for compatibility with standard microwave components, either as a single 50\,$\Omega$ for microstrip or two $100$\,$\Omega$ resistors in parallel for CPW line. The AuCu we deposit has a resistivity of 7.4 $\mu \Omega \text{cm}$, and so 25 squares of a 37 nm thick film are needed to realise a 50 $\Omega$ resistor, or 100 squares for the parallel arrangement.

\begin{figure}
\centering 
\includegraphics[width=14cm]{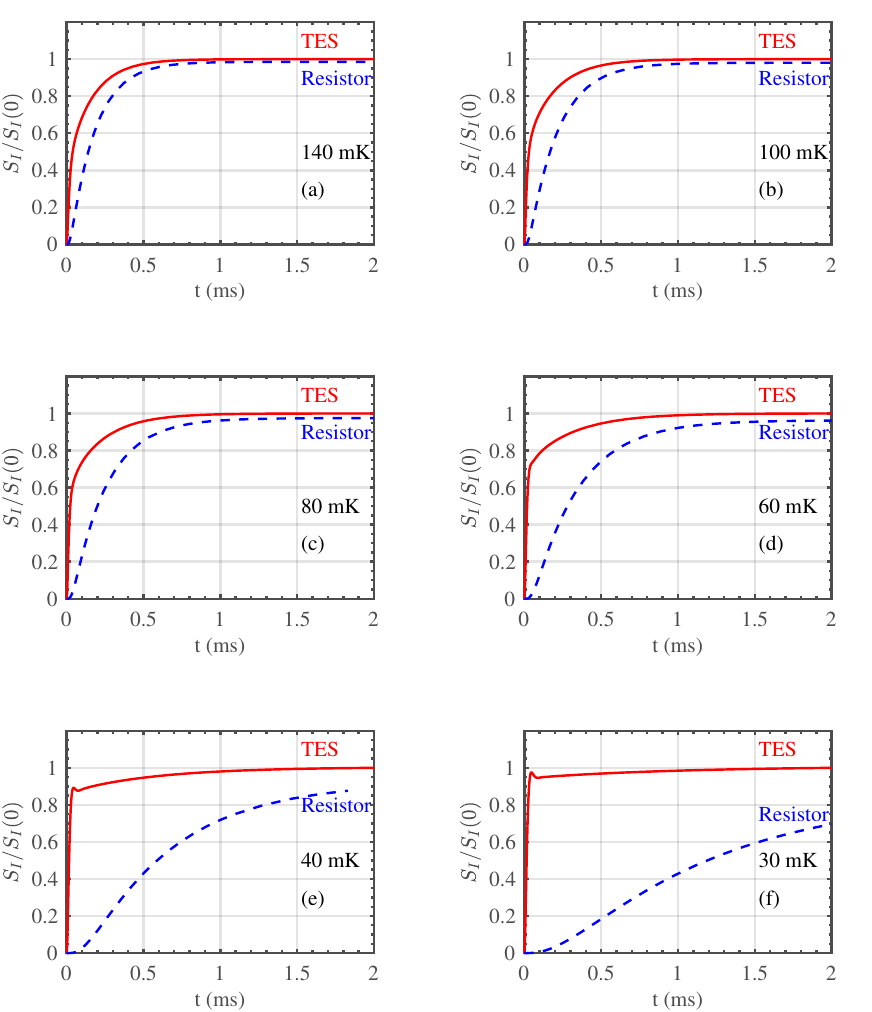}
\caption{\label{fig:Thermal_v6_times} Time-response  of the current per unit power-input $S_I$ in the TES or resistor for the distributed thermal model shown in Fig.~\ref{fig:Thermal_v6}.
Changes in current $\Delta I(t) $, for a step change in power $\Delta P$ absorbed at $t=0$ are plotted as $S_I(t)=\Delta I(t)/\Delta P$, normalized by   the DC power-to-current responsivity of the TES itself, $S_I(0)$ at each temperature.  Power is absorbed either in the TES (full red  lines) or in the resistor (dotted blue lines). Calculations are at six operating temperatures spanning $140$ to $30\,{\rm{mK}}$, 
plots (a) - (f), as indicated.}
\end{figure}

\subsubsection{Response of a distributed TES and separate load resistor on SiN}\label{TES_dist_load_response}

Figure~\ref{fig:Thermal_v6_times} shows the calculated response times of the output current of the TES $\Delta I(t) $  for a step change in power $\Delta P\vert_{t=0}$ occurring at time $t=0$ for both the TES and termination resistor. We show  the current-to-power response such that $S_{I}(t)=\Delta I(t)/\Delta P $ is the TES current response per unit power applied for a step change in absorbed power in the TES (full red lines) or in the termination resistor (dotted blue lines), all calculated for the distributed thermal model shown in Fig.~\ref{fig:Thermal_v6}.  In each instance, $S_{I}$ is further normalized by the DC responsivity of the TES $S_I(0)$ at the operating temperature - this approach highlights incomplete power detection as input power is lost to the heat bath due to the distributed nature of the thermal layout. In other words, power may flow off of the suspended island without increasing the temperature of the TES bilayer. Calculations are shown for six TES operating temperatures ranging from $140$ to $30\,{\rm{mK}}$, plots (a) - (f), respectively as indicated. The TES current response $S_I(t) $ with respect to power applied to the {\it termination} resistor becomes noticeably slower as the operating temperature is reduced. At the same time, the TES DC responsivity $S_I(0)$ i.e. $(t\to \infty)$ - not shown - increases, as expected, as  the operating temperature is lowered. 

For the TES bilayer itself, at $30\,{\rm mK}$ the current-power step response starts to show the onset of instability as the ETF approaches critical damping, again as expected. As the operating temperature increases, the initial fast response of the TES remains evident, but now enhanced thermal conduction from the TES to the adjacent heat capacities occurs and the current response shows an enhanced contribution from  a slower tail and the slight instability becomes increasingly less pronounced.  

By contrast, and for the same reasoning,  the TES current response to a step change in power absorbed in the RF absorbing resistor (i.e. the {\it RF detection  time-response} - shown in blue) becomes markedly {\it slower} as the operating temperature is reduced. The thermal coupling of the TES to the distributed heat capacities is reduced as the temperature is lowered, and the effect of ETF on overall detector response likewise is reduced. In Sec.~\ref{sec:TES_on_Si} we explore the origin of this slower response, and show that  electron-phonon decoupling in the absorbing resistor is the dominant source of the response's roll-of.

\begin{figure}
\centering
\includegraphics[width=14cm]{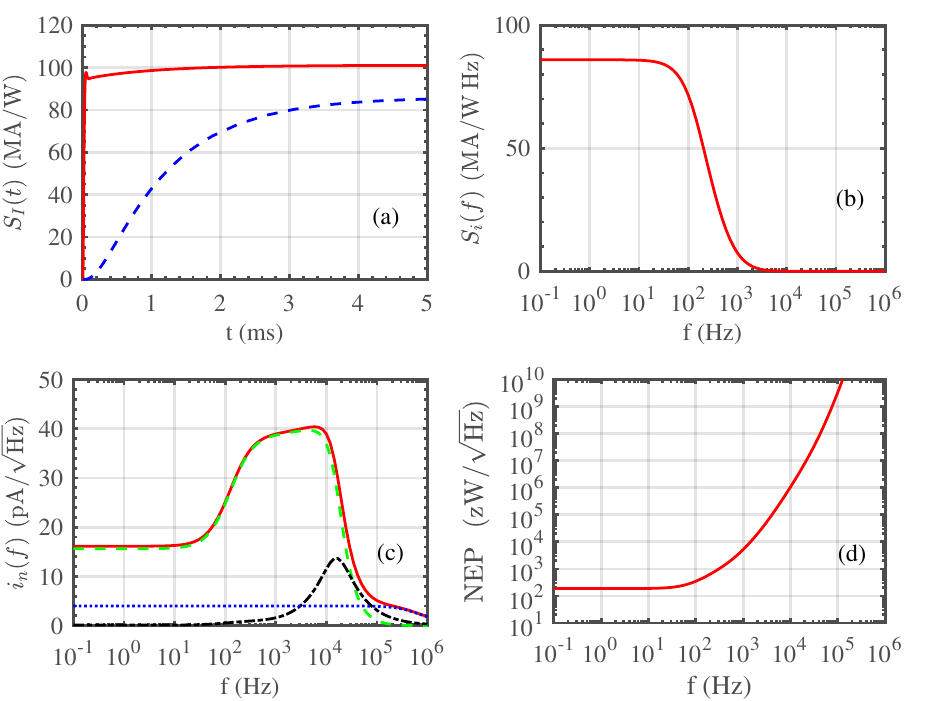}
\caption{\label{fig:TES_on_SiN} Response of a distributed TES detector on SiN operating at 30 mK. (a) Time response of the current in the  TES  for a step change  in power at $t=0$,   (red line) Power absorbed directly in the TES, or (blue dashed line) in the termination resistor. (b) Frequency response  for power absorption in the resistor. (c) Current noise spectral contributions: (green dashed) phonon noise, (black dash-dot) Johnson noise, (blue dot) SQUID readout noise and  (full red line) the total noise. (d) The calculated NEP.}
\end{figure}

Figure~\ref{fig:TES_on_SiN} further explores the performance  at $30\,{\rm mK}$ of a distributed TES and termination resistor on SiN. Fig.~\ref{fig:TES_on_SiN}~(a) shows the  time response of the current in the  TES  for a step change  in power at $t=0$. The  red line shows the effect of power absorption directly in the TES, the blue line for power absorption in the termination resistor. Fig.~\ref{fig:TES_on_SiN}~(b) shows the TES output frequency response  for power absorption in the termination resistor. Fig.~\ref{fig:TES_on_SiN}~(c) shows the calculated current noise spectral contributions: (green dash-dot) phonon noise, (black dashed) Johnson noise, (blue dashed) SQUID readout noise and red (full) the total noise, and (d) NEP.

\subsection{TES with stacked load resistor on SiN}\label{TES_stacked_load}

\begin{figure}
\centering
\includegraphics[width=12cm]{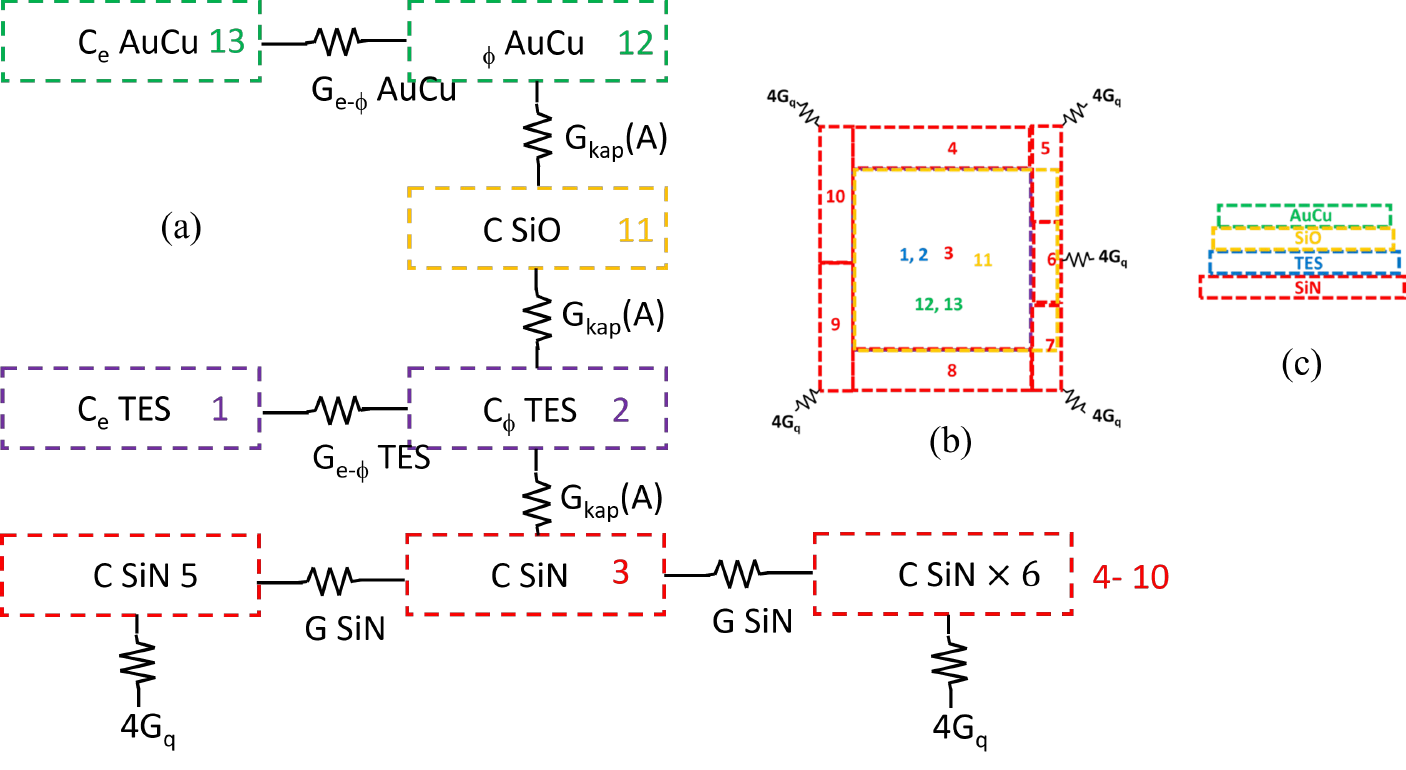}
\caption{\label{fig:stacked_TES_model} (a) Thermal model for a stacked TES and load resistor (b) plan view of the concept  and (c) vertical cross-section.}
\end{figure}

\begin{figure}
\centering
\includegraphics[width=14cm]{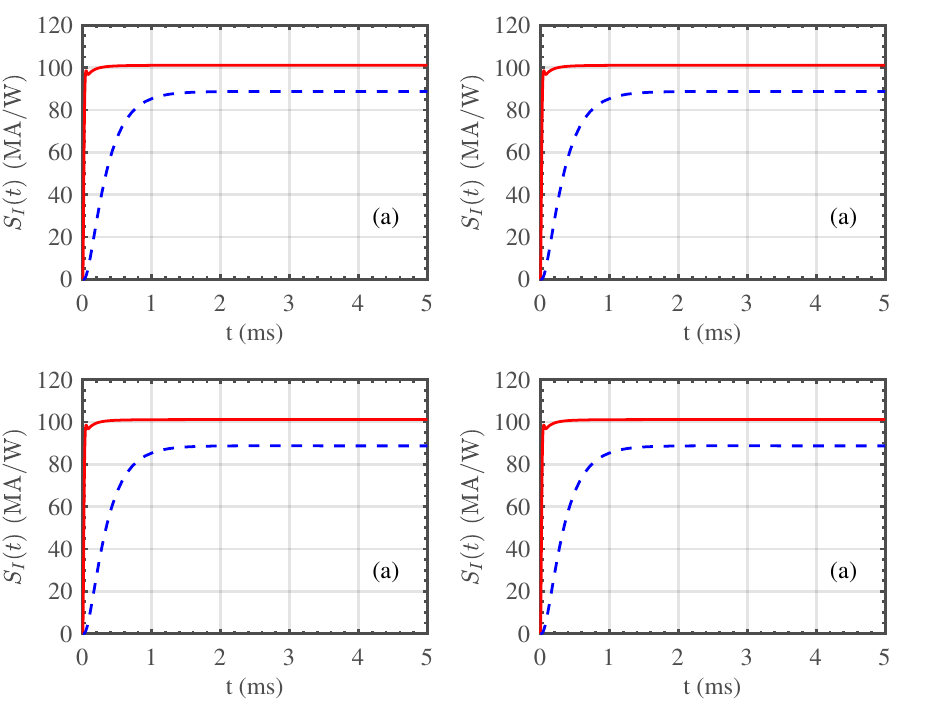}
\caption{\label{fig:stacked_TES_response} Response of a stacked TES on SiN  as shown schematically in Fig.~\ref{fig:stacked_TES_model}~(a) operating at 30 mK.
 (a) Time response of the current in the  TES  for a step change  in power at $t=0$, (red line) incident directly on the TES, or (blue dashed line) on the absorbing resistor. (b) TES responsivity with frequency for power absorption in the resistor. (c) TES current noise contributions: (green dashed) phonon contribution, black (dash-dot) Johnson noise, blue (dotted)  SQUID readout  and, red (full) the total noise. (d) The calculated NEP.}
\end{figure}

Figure~\ref{fig:stacked_TES_model} shows a thermal model for a TES where  the power-absorbing resistor is formed directly above the TES itself, but electrically isolated from the TES by an additional silicon dioxide layer (assumed to be $400\,{\rm nm}$ thick). Here we explicitly include the Kapitza conductance between the thin films - the dielectric films are sufficiently thin to ensure that the transverse thermal conductance is much higher than $G_{K}$. 

\subsubsection{Response of a stacked TES and Resistor on SiN}\label{Stacked_TES__response}

Figure~\ref{fig:stacked_TES_response} shows the  response of the stacked TES on SiN  shown schematically in Fig.~\ref{fig:stacked_TES_model}~(a) operating at 30 mK. (a) shows the time response of the current in the  TES  for a step change  in power at $t=0$ (red line) incident directly on the TES, and (blue line) in the absorbing resistor. (b) shows the current responsivity as a function of frequency for power absorption in the resistor, (c) shows the TES current noise spectrum and contributions ((green dash-dot phonon contribution, black dashed Johnson noise, blue dashed SQUID readout and red full ,total) and (d) calculated NEP.
 
The TES's response to direct power absorption (Fig.~\ref{fig:stacked_TES_response}(a) red line) is almost identical to the response calculated for the distributed model (Fig.~\ref{fig:TES_on_SiN}(a) red line). The response remains mainly characteristic of the TES and the thermal link to the bath as these are unchanged from the distributed model of Sec.~\ref{TES_dist_load}. However, the response of the TES to power absorbed in the RF termination resistor is noticeably different (compare the blue lines in Fig.~\ref{fig:stacked_TES_response}(a) to Fig.~\ref{fig:TES_on_SiN}(a)), or in the frequency domain compare Fig.~\ref{fig:stacked_TES_response}(b) to Fig.~\ref{fig:TES_on_SiN}(b). The distributed detector response is limited by heat diffusion in SiN. The stacked geometry is becoming limited by electron-phonon decoupling in the termination resistor and has wider usable readout bandwidth. We compare response times quantitatively for all models later in Sec.~\ref{sec:Compare_TES_response_times}. 

\subsection{Distributed TES on Silicon}\label{sec:TES_on_Si}
\begin{figure}
\centering
\includegraphics[width=14cm]{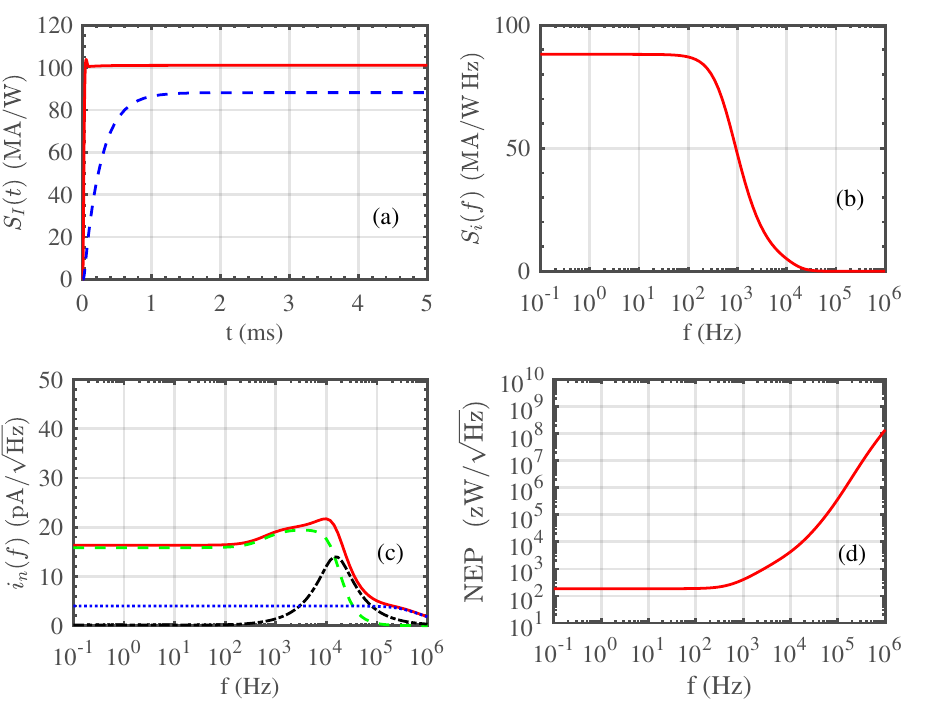}
\caption{\label{fig:TES_on_Si} Response of  a TES on Si operating at 30 mK. (a) Time response of the current in the  TES  for a step change  in power at $t=0$,  (red line) incident directly on the TES, or (blue line) on the absorbing resistor. (b) Responsivity with frequency for power absorption in the resistor. (c) TES current noise contributions: (green dashed) phonon contribution, black (dash-dot) Johnson noise, blue (dotted)  SQUID readout, and red (full) the total noise. (d) The calculated NEP.}
\end{figure}
\begin{figure}
\centering
\includegraphics[width=8cm]{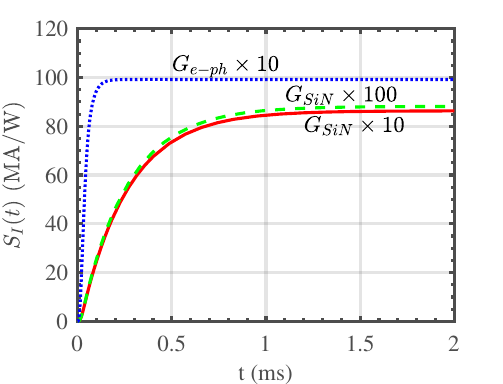}
\caption{\label{fig:TES_on_Si_Geph} Response risetimes for a step change in power absorbed in the resistor of  a distributed TES with separate absorbing resistor fabricated on Si. Red (full) and (dashed) green lines: effect of increasing the value used for the Si conductance by a factor of 10 and 100  respectively. (Dotted) blue line: the  effect of increasing $G_{e-ph}$ of the resistor by a factor of $10$. The TES operates at $30\,{\rm mK}$.}
\end{figure}

In this section, we consider the operation of a distributed TES on crystalline Si. For the heat capacity we assume that the Si can be described using the results of the Debye model described in  Sec~\ref{sec:C_dielectric}. We expect that thermal transport is limited by phonon scattering off imperfections on the boundaries of the sample rather than being diffusively scattered in bulk \cite{Asheghi_1998}. We have not at this stage attempted to fully model heat transport in  the crystalline Si substrate, but rather for the majority of the modelling we simply increased the (diffusive) values used to describe SiN by factors of 10 and 100. The results show that heat transport in the substrate is not in any case a limiting factor in overall performance at the lowest temperatures. Note that the heat transport in the legs is already ballistic, and so the earlier descriptions still apply, resulting in each legs contributing 4 acoustic modes at low temperature. TESs fabricated on Si have been demonstrated in practice and deployed on experiments \cite{rostem2014precision,rostem2016silicon}.

\subsubsection{Response of a distributed  TES and resistor on Si}\label{Distributed_TES_Si_response}

Figure~\ref{fig:TES_on_Si_Geph} shows the predicted response for a step change in power absorbed in the resistor of  a distributed TES and resistor fabricated on Si. The red and dashed green lines show the effect of increasing the conductance by (based on that of SiN) by factors of 10 and 100  respectively. The dash-dot blue line shows the effect of increasing $G_{e-ph}$ of the resistor by a factor of $10$. It can be seen that heat flow in the substrate is not dominant in limiting the overall response time of a TES on crystalline Si. The response is limited by electron-phonon decoupling in the termination resistor and TES. 

\subsection{Comparison of response times for the TES models} \label{sec:Compare_TES_response_times}

\begin{figure}
\centering
\includegraphics[width=8cm]{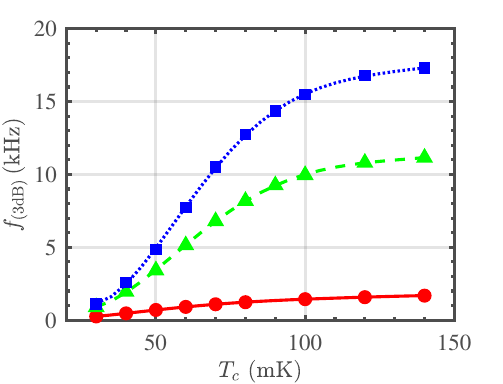}
\caption{\label{fig:3_models_f3bB} Comparison of the predicted 3-dB-roll-off frequencies for the 3 TES models as a function of operating temperature $T_c$. Red circles distributed on SiN, green triangles stacked geometry, and blue squares distributed on Si. Solid lines are a guide to the eye.}
\end{figure}
\begin{table}
\centering
\begin{tabular}{lcc}
\toprule
\textbf{Model} & $f_{{\rm 3-dB}}$& $\Delta B$  \\
    &  ${\rm Hz}$ & ${\rm Hz}$\\
\midrule
 Distributed on SiN & 250 & 64\\
Stacked on SiN &890 & 315 \\
 Distributed on Si & 1100 &  520\\
\bottomrule
\end{tabular}
\caption{\label{tab:TES_f3dB}
3-dB roll-off frequencies for the 3 TES models studied here at $30\,{\rm mK}$.
}
\end{table}

Here we quantitatively compare the response times of the 3 TES models described in Secs.~\ref{TES_dist_load} to \ref{sec:TES_on_Si}. Figure~\ref{fig:3_models_f3bB} shows the calculated 3-dB roll-off of the modelled power-step responsivity as a function of $T_c$ for the  3 models. The red circles and line show  results for a distributed detector on SiN, the green triangles and line for a stacked detector on SiN and the blue squares and line for a distributed detector on Si. In all cases lines are a polynomial fit to the data and are included to guide the eye. 

Table~\ref{tab:TES_f3dB} shows the calculated 3-dB roll-off frequencies for the TES responsivity for detection of a step change in power in the resistor using the 3 TES models described above at $30\,{\rm mK}$. Entries labelled $\Delta B$ show the useable 3-dB roll-off frequencies calculated from the calculated  NEP i.e. including all modelled additional noise sources.  Note that these additional noise sources include thermal exchange noise between components. The faster heat transport for a TES on Si moves the excess noise peak to higher frequencies and $\Delta B$ increases. The response bandwidths of the stacked TES on SiN and distributed TES on Si are comparable at $30\,{\rm mK}$: in these cases electron-phonon decoupling in the absorbing resistor is the main factor determining the bandwidth. For the distributed TES on SiN, heat diffusion in the SiN  further reduces the both the response and useable baseband bandwidth.  

\subsection{Thermal Modelling of  electron-phonon decoupled  nano-HEB  at 30 mK  }\label{sec:HEB_performance}

\begin{figure}[h!]
\centering
\includegraphics[width=4cm]{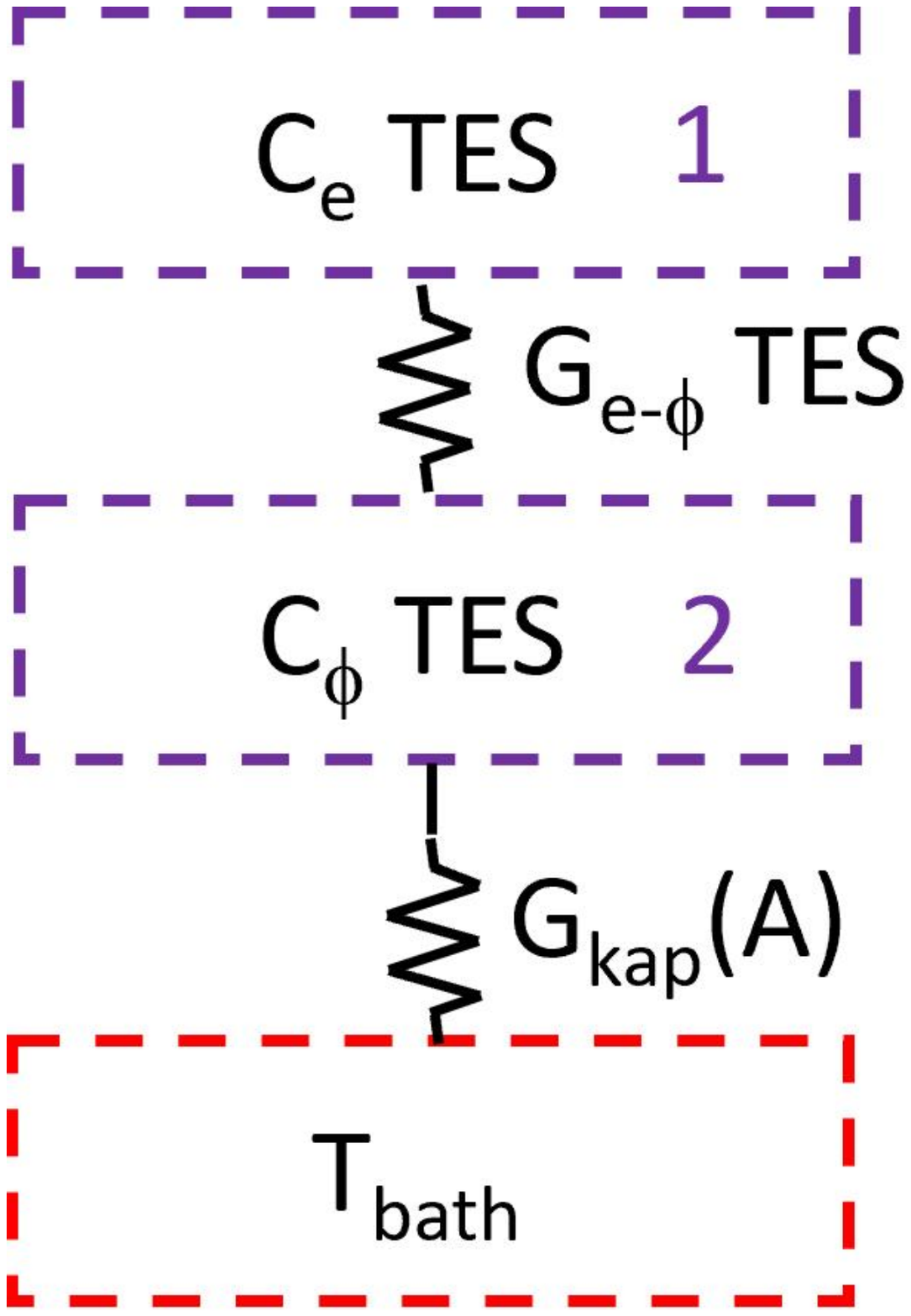}
\caption{\label{fig:Simple_HEB} Thermal model for a simple nano-HEB with electron-phonon decoupling. We assume that the nano-HEB itself is the power absorbing element.}
\end{figure}

In this section, we explore the performance of a nano-HEB operating at $30\,{\rm mK}$. In the context of the modelling carried out here, this device is essentially a TES relying on electron-phonon decoupling, as described in Ref.~\cite{cabrera1998detection}, but with a lower operating temperature. We used the  parameters of Tab.~\ref{tab:HEB concept} as inputs to the modelling. Figure~\ref{fig:Simple_HEB} shows the thermal model used. We assume that power is absorbed directly by the electron system of the nano-HEB. 

\begin{figure}
\centering
\includegraphics[width=14cm]{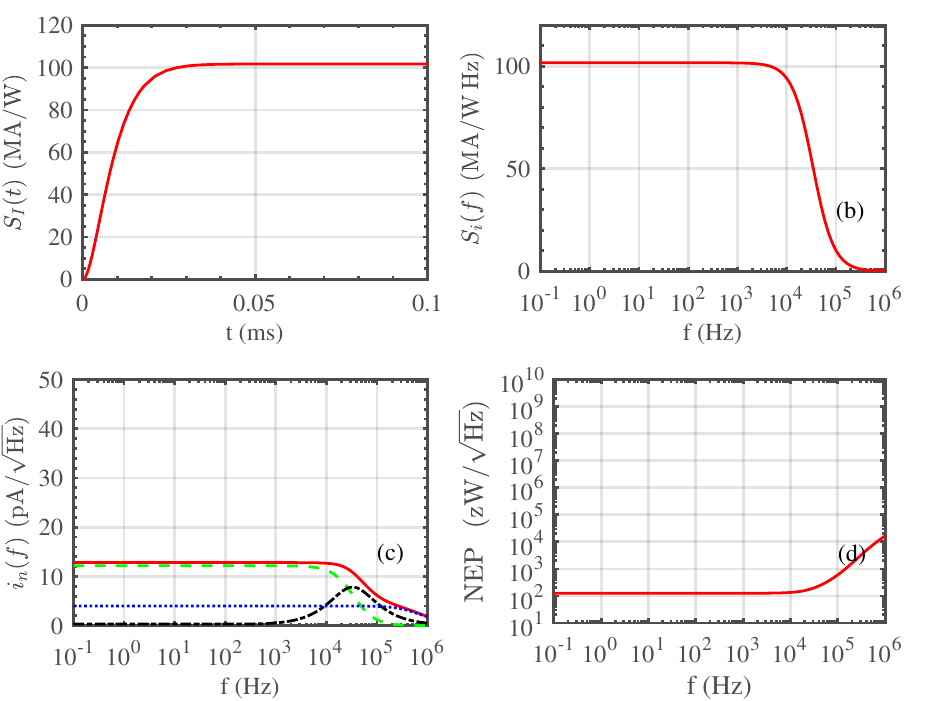}
\caption{\label{fig:Simple_HEB_results} Results for a simple nano-HEB operating at 30 mK. (a) Time response for a step  change in power incident on the nano-HEB. (b) Responsivity with frequency. (c) nano-HEB current noise contributions: (green dashed) nano-HEB contribution, black (dash-dot) Johnson noise, blue (dotted)  SQUID readout, and red (full) the total noise. (d) the calculated NEP.}
\end{figure}

When modelling the performance at $30\,{\rm mK}$, we found that using the nano-HEB electrothermal parameters 
described in Sec.~\ref{sec:tes_designs} resulted in significant response ringing at short-times, and so a reduced value of $\alpha=100$ was used to ensure stability. Figure~\ref{fig:Simple_HEB_results} shows results of the modelling. Figure~\ref{fig:Simple_HEB_results}(a) shows the time response of the nano-HEB current for power absorbed directly in the detector. Note the change in scale on the x-axis compared to previous figures. Figure~\ref{fig:Simple_HEB_results}(b)  shows the frequency response of the  nano-HEB current for direct power absorption. Figure~\ref{fig:Simple_HEB_results}(c) shows the total  nano-HEB current noise (solid red line) and contributions.  There is very little  contribution from  phonon exchange noise - the green dash-dot line can be described very closely by  a single pole roll-off. 
 
The  3-dB roll-off of the responsivity in  Fig.~\ref{fig:Simple_HEB_results}(b) occurs at $34.9\,{\rm kHz}$. This is significantly faster than the frequencies shown in Table~\ref{tab:TES_f3dB}. For a  nano-HEB, the response  roll-off is determined by  electron-phonon decoupling in the nano-HEB, but in this case electrothermal feedback further increases the baseband bandwidth. 

\subsubsection{RF Coupling to a low resistance  nano-HEB}\label{sec:HEB_coupling}

Although  nano-HEBs have certain attractive features in the context of realising microwave and millimetre-wave detectors,  a key question remains about whether a {\it low-resistance} absorber, with $R_n\sim 200\,{\rm m\Omega}$, suitable for reading out with conventional multi-stage SQUIDs, can be efficiently coupled to the RF components needed for making a homodyne system. Ideally, the RF components, such as the power-splitter/pump-injector, should be fabricated in thin-film on the same substrate as the two balanced detectors. Superconducting microstrip having $Z_0 <$ 20 $\Omega$ can be fabricated easily, but a higher $Z_0$ needs more extreme geometries. Because the integrated RF components must be coupled to the external RF source, $Z_0 =$ 15 $\Omega$ is a suitable choice, but this is well away from the input resistance of the HEB. To achieve a power match, additional impedance transformers are needed, but because the impedance ratio is so high (50 to 100) optimum behaviour can only be achieved over narrow bandwidths. At short wavelengths, another approach is to use a low source impedance. For example, whereas an electric field probe in a cavity is an intrinsically high-impedance device,  a magnetic loop is intrinsically a low-impedance device. Thus in principle, the RF impedance of the system could be just a few $\Omega$'s, which could be matched more easily to the   $R_n\sim 200\,{\rm m\Omega}$ input impedance of the HEB. We conclude that although the very low impedance of an HEB-based homodyne detector creates difficulties, these are not, in principal at least, insurmountable.  

\subsubsection{Microstrip-impedance matched  nano-HEB}\label{sec:Impedance_matched _HEB}

It is interesting to consider briefly whether a higher resistance  nano-HEB (with $R_n\sim 20\,{\rm \Omega}$ (sufficient to be coupled more easily to typical CPW or microstrip line impedances) could be readout using conventional Si-based electronics. Assuming $T_c=30\,{\rm mK}$, $R_0\sim 5 \,{\rm \Omega}$, we estimate a  Johnson current noise of order $i_J=0.60\,{\rm pA/\sqrt{Hz}}$. In order to be competitive, conventional readout would require current noise below $i_J$ 
and voltage noise $v_n<i_J R_0 \sim 3\,{\rm pV/\sqrt{Hz}}$. Such performance is not obviously easily available even for cooled Si-based electronics, and so the RF matching problem remains problematic. 

\section{Multilayer $T_c$ modelling and measurement}\label{sec:Tc_model_measure}

Consider how a TES, or indeed pair-breaking superconducting detector, could be engineered for operation at very low temperatures $<$ 50 mK. We base our modelling on the superconducting proximity effect in normal-superconducting (S-N) bilayers, which arises from a low-order solution of the Usadel equations as described by Martinis {\it et al.}\cite{martinis2000calculation} for thin-films, and extended by Zhao {\it et al.}\cite{zhao2018calculation} to multilayers (S-N-S, N-S-N, etc.).

\subsection{$T_c$ calculation in SN bilayers}\label{sec:Tc_model}

Martinis {\it et al.}\cite{martinis2000calculation} describe a calculation of $T_c$ for thin-film S-N bilayers:
\begin{equation}\label{eqn:Martinis}
\ln\frac {T_c} {T_{c0}}= - \int^{h \omega_D}_0 \frac {dE}{E} \left[ \frac {d_N N(0)_N} {{d_N N(0)_N+d_s N(0)_S}} 
\frac{ 1}  {1+ E^2/\tau^2} \right] \tanh { \frac {E} {2k_b T_c}},
\end{equation}
where $d_N$, $d_S$ are the thicknesses of the $N$ and $S$ layers, $N(0)_N$, $N(0)_S$ are the single-spin densities of states for electrons at the Fermi surface in $N$ and $S$ respectively, $\omega_D$ is the Debye frequency and $h$ is Planck's constant. $T_{c0}$ is the transition temperature of $S$ in the absence of $N$. $\tau$ is a parameter accounting for the energy coupling between $N$ and $S$. Martinis {\it et al.} show that
\begin{equation}\label{eqn:tau_prox}
\tau=\frac {2t} {\pi \lambda_F^2} \left(  \frac {1} {d_N N(0)_N} +\frac {1}{d_S N(0)_S}\right),
\end{equation}
where $\lambda_F$ is the Fermi wavelength (here we assume equal to the Fermi wavelength of Au) and $t$ is a unitless interface transmission coefficient. Martinis {\it et al.} suggest  we should expect $t\sim 0.1$ for most clean interfaces, but in practice we hafve found this to vary significantly depending on the precise deposition conditions. 

\subsection{Multilayer modelling}\label{sec:Multilayer_model}

For general multilayers  the key point, as demonstrated by Zhao, is that a thin-film multilayer must have a common transition temperature for all components. An ${\rm N_1 - S - N_2} $ multilayer for example, can then be  treated as two bilayers the first of thicknesses $d_{N1}$, $d_x$ and a  second bilayer of thicknesses $(d_S - d_x),  d_{N2}$.The  problem amounts to finding a length $0<d_{x}<d_S$, such that Eqs.~\ref{eqn:Martinis} and \ref{eqn:tau_prox} are satisfied with a common $T_c$ across the entire film. In practice finding $d_x$ is numerically straight-forward. The advantage of this approach is that we can also explore separate values for the interface transmission coefficients $t_1$ and $t_2$ between $N_1 - S$ and $S - N_2$.

\subsection{Designing a multilayer for $30\,{\rm mK}$ operation}\label{Multilayer_approach}
\begin{table}
\centering
\begin{tabular}{lccccc}
\toprule
\textbf{Reference} & $d_{\rm Ti}$ & $d_{\rm Au}$ & $d_{\rm Ti}$ & $T_c $ & $t$ \\
    				    &  ${\rm nm}$ & ${\rm nm}   $   & ${\rm nm} $  & ${\rm mK}$ & \\
\midrule
Bilayer 1 & 55  & 240  &  & 225 &  0.075 \\
Bilayer 2 & 40  & 240  &  & 136 &  0.075 \\
Bilayer 3 & 33  & 240  &  & $<$85 &  $>$0.077 \\
Bilayer 4 & 35  & 240  &  & 104 &  0.075 \\
Bilayer 5 & 40  & 240  &  & 153 &  0.07 \\
Bilayer 6 & 40  & 240  &  & 140 &  0.075 \\
Bilayer 7 & 65  & 200  &  & 328 &  0.06 \\
Bilayer 8 & 40 & 240  &  & 129 &  0.079 \\
Bilayer 9 & 40 & 240  &  & 129 &  0.063 \\
Bilayer 10 &   & 20  & 100 & 403&  $\sim 6$ \\
\bottomrule
\end{tabular}
\caption{\label{tab:Bilayer_Tc}
Modelled proximity interface transmission coefficients for bilayers.
}
\end{table}

TiAu bilayers have already been reported widely. Our experience and also that of the South Pole Project (SPT)\cite{carter2018tuning,ding2016optimization,yefremenko2018impact} is that thin-film Ti has unexpected metallurgical reactions with some widely-used higher $T_c$ superconductors - particularly Nb, that is often chosen for electrical interconnects. The interaction may-well be associated with hydrogen diffusion along grain boundaries leading to changes in  electrical and morphological characteristics and a resistive contact where in reality zero resistance is required. 

One solution to  the problem, which we have already demonstrated, is to implement metallurgically robust, $T_c$-graduated interconnects (such as Nb-Al-Ti) but this approach increases the overall footprint of the complete detector, probably leading to  increased heat capacity hence reduced output bandwidth, and also increased processing complexity. The approach we have explored here was whether an encapsulated Au-Ti-Au multilayer could be implemented. The SPT project have also reported that a thin ($5\,{\rm nm}$)  Ti ``seed'' layer that was deposited first on the SiN producing a quadlayer of (Ti)-Au-Ti-Au, gave improved reproducibility of measured multilayer $T_c$. 

\subsection{Measurement of bilayer and multilayer $T_c$}\label{Multilayer_measurement}

Transition temperatures of single Ti films, Ti-Au bilayers and multilayers were measured using a refrigerator with base temperature of $90\,{\rm mK}$. For Ti films  deposited on SiN we found a close-to linear dependence on thickness between $d_{Ti}=25\,{\rm nm}$ with  $T_{c0}=470\,{\rm mK}$ and $d_{Ti}=100\,{\rm nm}$ with $T_{c0}=550\,{\rm mK}$. We have included this dependence when using Eqs.~\ref{eqn:Martinis} and \ref{eqn:tau_prox} in modelling.
 
\subsubsection{Bilayer $T_c$}\label{bilayer_Tc}

Table~\ref{tab:Bilayer_Tc} shows measured $T_c$ values at typically $1$ or $3\,{\rm \mu A}$ for ten calibration TiAu (or AuTi) bilayers. The depositions are listed by date order covering 18 months. For bilayers 1 to 9 where the Ti layer was deposited first, estimated transition coefficients are very consistent and we find $t=0.72  \pm 0.01$. For bilayer 10, where the Au was deposited first we find $t\sim 6$. This is significantly different from the TiAu measurements and  strongly suggests that the model or more likely one of its assumptions does not apply in this case. 

\begin{table}
\centering
\begin{tabular}{lccccccc}
\toprule
\textbf{Reference} & $d_{\rm Ti}$ & $d_{\rm Au}$ & $d_{\rm Ti}$ &  $d_{\rm Au}$ & $T_c $ & $t_1$ & $t_2$ \\
    				    &  ${\rm nm}$ & ${\rm nm}   $   & ${\rm nm} $    & ${\rm nm} $ & ${\rm mK}$ &  &\\
\midrule
Multilayer 1 & 5  & 20  & 50 & 205 &  130 & 0.07  & 0.072 \\
Multilayer 2 & 5  & 20  & 55 & 225 &  142 & 0.11  & 0.072 \\
Multilayer 3 & 5  & 20  & 46 & 225 &  $<87$ & $>0.25$  & 0.072 \\
Multilayer 4 & 5  & 20  & 50 & 225 &  $<87$ & $>1.0 $ & 0.072 \\
Multilayer 5 & 5  & 20  & 55 & 225 &  138 & 0.13  & 0.072 \\
Multilayer 6 &   & 30  & 65 & 220 &  $<100$  & $>1.0 $  & 0.072 \\
Multilayer 7 &   & 30  & 55 & 220 &  $<100$  & $>1.0 $  & 0.072 \\
Multilayer 8 &   &220 & 65 &30&  $<90$  & $>8 $  & 0.072 \\
Multilayer 9 &   &20 & 80 &220&  $<90$  & $>10 $  & 0.072 \\
Multilayer 10 &  5 &20 & 55 &225&  150 & 0.08  & 0.072 \\
Multilayer 11 &  5 &20 & 55 &225&  150 & 0.08  & 0.072 \\
Multilayer 12 &    &20 & 80 &220&  150 & $>10 $  & 0.072 \\
Multilayer 13 ({\bf Si}) &    &20 & 80 &220&  241 & 0.32 & 0.072 \\
\bottomrule
\end{tabular}
\caption{\label{tab:Multilayer_Tc}
Modelled proximity interface transmission coefficients for Multilayers. We assume that for the upper Ti-Au interface transmission coefficient $t_2=0.072$ the mean value from Table~\ref{tab:Bilayer_Tc}. Note all depositions were on SiN except Multilayer 13 that was on Si.
}
\end{table}
\subsubsection{Multilayer $T_c$}\label{Multilayer_Tc}

Table~\ref{tab:Multilayer_Tc} shows measured $T_c$ for multilayers. Depositions 1 to 12 used SiN substrates, Deposition 13 was on Si.  For the upper Ti-Au interface we assume $t_2=0.072$ the mean value determined from Table~\ref{tab:Bilayer_Tc}. For quad layers having a $5\,{\rm nm}$ Ti seed layer (Multilayers $1-5$, 10, 11) we ignore the effect of the thin Ti. 

For those Ti-seeded multilayers with a measurable $T_c$ we find $t_1=0.098\pm 0.028$ to describe the Au-Ti interface. This is close to the value for Ti-Au bilayers given the measured variance. 
The seeded multilayers without a measurable transition to $\sim 90 \,{\rm mK}$ appear to be anomalous. 

For multilayers without the $5\,{\rm nm}$ Ti seed, $t_1$ becomes large as for the Au-Ti bilayer reported in Table~\ref{tab:Bilayer_Tc}. A similar explanation is possible: assumptions of the modelling appear to breakdown for Ti deposited directly onto Au on SiN, possibly due to morphology changes in the Ti, perhaps due to changed film granularity and/or low intrinsic $T_{c0}$ for Ti films on Au.

We conclude that a fully encapsulated Au-Ti-Au multilayer may not be feasible (or at least very difficult to calibrate and certainly difficult to predict). A quad-layer with a thin Ti seed would be possible and also encapsulate the main Ti layer giving some resilience to  subsequent ageing effects for example. This solution would require $T_c$-graded leads with additional processing steps.  A (Ti)-Au-Ti-Au multilayer on Si also appears possible. 
\subsection{Recommendation for a $30\,{\rm mK}$ detector}\label{Layup}
We assume a quad-layer of Ti-Au-Ti-Au with $5\,{\rm nm}$ Ti seed layer, $t_1=0.1$, $t_2=0.072$.
We aim for a sheet resistance $R_{sq}=60\,{\rm m\Omega}$, with film thicknesses such that $T_c=30\,{\rm mK}$. 
\begin{figure}
\centering
\includegraphics[width=8cm]{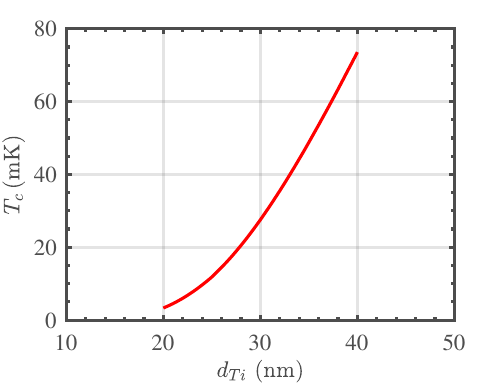}
\caption{\label{fig:Tc_model} Calculated (Ti)-Au-Ti-Au multilayer $T_c$ as a function of Ti thickness for a $5-20-d_{Ti}-200\,{\rm  nm}$ multilayer
based on measurements shown in Table~\ref{tab:Multilayer_Tc}.
}
\end{figure}
Figure~\ref{fig:Tc_model} shows calculated  (Ti)-Au-Ti-Au multilayer $T_c$ as a function of Ti thickness for a $(5)-20-d_{Ti}-200\,{\rm  nm}$ multilayer. Sheet resistance is calculated to be $R_{sq}=59\,{\rm m\Omega}$. A Ti thickness of $32\pm1\,{\rm nm}$ is predicted to show $T_c=35\pm 5 \,{\rm mK}$. 
Such a multilayer would be suitable for both a TES or  nano-HEB for this transition temperature.

\section{Summary and Conclusions}\label{sec:Conclusions}

In a companion paper \cite{Chris2023}, we presented an extensive theoretical study of the feasibility and likely performance of microwave and millimetre-wave homodyne detectors based on ultra-low-noise superconducting power detectors. This theoretical study showed that excellent quantum-noise-limited performance would be achieved over the whole of the 1 GHz - 100 GHz range if a power detector could be identified that (i) can operate at a base temperature approaching 10 mK, (ii) that has a threshold NEP that allows the quantum noise temperature limit to be reached at microwave frequencies, (iii) that has, simultaneously, a saturation power that is sufficient to accommodate a pump signal that is at least a few times higher than the actual signal, and (iv) whose noise performance is not degraded, through say heating, when the pump is applied. In this report, we have considered extensively whether such a device exists, and whether the development of an integrated homodyne microwave technology is within the reach of a modest technology development programme.

General findings are as follows:
\begin{enumerate}

\item{\it All superconducting bolometric detectors} operating close-to the thermodynamic limit at 20-50 mK are expected to have comparable performance. There is nothing intrinsic about the solid-state physics of TESs and HEBs that prevents ultra-low-noise operation, and the realisation of integrated homodyne detectors achieving the quantum noise temperature at microwave frequencies and above. From an applications perspective, this would be an extraordinary device, and could make a substantial contribution to haloscope-type searches for ultra-light dark matter. 

\item Although such a device has not yet been demonstrated experimentally, we feel that it is well within the grasp of current design and device manufacturing techniques, with both TESs and certain HEBs being of particularly interest.

\item TESs with ballistic thermal isolation operating at low temperature $\sim 30\,{\rm mK}$ are a very promising and mature technology for this application. Although ultra-low-noise TES technology is a widely used technology, pushing the $T_{c}$  down from $\sim$100 mK to $<$30 mK requires experimental work.  We have already demonstrated TESs with ballistic support legs, \cite{osman2014transition}, and found these to be highly reproducible, and have given excellent performance. We are confident of being able to manufacture small arrays of balanced pairs straightforwardly. These are read out using conventional DC SQUIDs, and so no additional microwave readout technology is needed.

\item A TES operated at $T_c=30\,{\rm mK}$ with ballistic phonon-transport thermal isolation, having 
${\rm NEP}= 145\,{\rm {zW/\sqrt{Hz}} } $, would realise a homodyne detector having quantum-limited noise at frequencies above $5\,{\rm GHz}$.

\item We have identified a S-N multilayer suitable for both a TES or a nano-HEB using thin Ti and Au films on SiN for operation at $30\,{\rm mK}$.

\item A stacked TES geometry is expected to offer enhanced readout bandwidth compared with distributed designs, but the stacked geometry presents processing challenges.

\item An electron-phonon decoupled  nano-HEB may also be suitable whilst offering enhanced readout bandwidth. However, it is likely to be difficult to achieve efficient RF coupling to this necessarily low impedance device, particularly if a large tuning range $\nu_s=1-20\,{\rm GHz}$ is required.

\item Other  bolometers may be capable of operating at these temperatures with comparable performance, although significant development would be required, particularly those that are based on their own RF modulation schemes (SNS bolometers or KIDs) where the modulation frequency is typically close to the {\it signal} frequencies $\nu_s$ of interest. 

\end{enumerate}

Our own preference is to realise and demonstrate integrated homodyne detectors based on TESs of the kind shown in Figs \ref{fig:TES_concept} and \ref{fig:Mask_v1}, as we believe that these could make a unique contribution to a variety of fundamental physics experiments, particularly those based on reading out microwave cavities in the quantum ground state. A TES homodyne detector, together with digital band defining filters on the output, would provide a quantum noise limited device, requiring only baseband readout, that could be swept over a very large frequency range. There is nothing that limits, intrinsically, the operating frequency range of a microstrip, coplanar, or indeed discrete TES. In fact, we have operated these devices in other experiments at frequencies ranging from 10 GHz to 700 GHz, although above 100 GHz other technologies start to compete.

\bibliographystyle{unsrtnat}
\newcommand*{\doi}[1]{\href{http://dx.doi.org/#1}{doi: #1}}
\bibliography{references_dg}

\end{document}